\documentclass[a4paper,12pt]{article}
\usepackage[top=2.5cm, bottom=2.5cm, left=2.5cm, right=2.5cm]{geometry}
\usepackage{authblk, graphicx, amsmath, amssymb, booktabs, multicol, subcaption, multirow}

\title{Measurements and modelling of stray magnetic fields and the simulation of their impact on the Compact Linear Collider at 380 GeV}
\author[1,2]{C. Gohil\footnote{chetan.gohil@physics.ox.ac.uk}}
\author[1]{P.\,N. Burrows}
\author[2]{N. Blaskovic Kraljevic\footnote{Present address: European Spallation Source, Lund, Sweden.}}
\author[2]{D. Schulte}
\author[3]{B. Heilig}
\affil[1]{\small John Adams Institute (JAI), University of Oxford, Oxford, OX1 3RH, United Kingdom}
\affil[2]{\small The European Organization for Nuclear Research (CERN), Geneva 23, CH-1211, Switzerland}
\affil[3]{\small Mining and Geological Survey of Hungary (MBFSZ), Budapest 1590, Pf. 95, Hungary}
\date{\small (\today)}

\begin{document}
\maketitle

\begin{abstract}
The Compact Linear Collider (CLIC) targets a nanometre beam size at the collision point. Realising this beam size requires the generation and transport of ultra-low emittance beams. Dynamic imperfections can deflect the colliding beams, leading to a collision with a relative offset. They can also degrade the emittance of each beam. Both of these effects can significantly impact the luminosity of CLIC. In this paper, we examine a newly considered dynamic imperfection: stray magnetic fields. Measurements of stray magnetic fields in the Large Hadron Collider tunnel are presented and used to develop a statistical model that can be used to realistically generate stray magnetic fields in simulations. The model is used in integrated simulations of CLIC at 380\,GeV including mitigation systems for stray magnetic fields to evaluate their impact on luminosity.
\end{abstract}

\section{Introduction}
The Compact Linear Collider (CLIC)~\cite{clic-pip, clic-cdr} is an $e^+ e^-$ collider with three centre-of-mass energy stages: 380\,GeV, 1.5\,TeV and 3\,TeV. In this paper, we derive a model for stray magnetic fields and apply it to the 380\,GeV stage of CLIC.

\subsection{Dynamic Imperfections}
Dynamic imperfections in a linear collider lead to luminosity loss by deflecting the colliding beams and causing emittance growth. The nanometre beam size at collision in CLIC makes it particular sensitive to dynamic imperfections.

Beams in linear colliders are generated in pulses. Dynamic imperfections influence consecutive pulses differently. This makes them difficult to correct.  The main tool for mitigating the impact of dynamic imperfections is a beam-based feedback system, which measures and corrects the beam offset. Often the beam-based feedback system, whose bandwidth is inherently limited by the beam repetition frequency, is not enough to mitigate the imperfection to the desired level. Dedicated studies are necessary to devise a mitigation strategy for dynamic imperfections. In this paper, we look at the impact of stray magnetic fields and their mitigation.

\subsection{Stray Fields}
Stray magnetic fields, or simply stray fields, are external dynamic magnetic fields, which influence the beam. They can be classified in terms of their source: natural, environmental and technical~\cite{sf-measurements, sf-impact, edu}.

\subsubsection{Sources}
\textbf{Natural} stray fields are from non-man-made objects, e.g. the Earth's magnetic field. A review of natural stray fields can be found in~\cite{balazs}. Natural stray fields have large amplitudes at low frequencies, which can be mitigated with a beam-based feedback system~\cite{integrated-simulations, lumi-paper}. At higher frequencies the amplitude is small enough that they can be ignored.

\textbf{Environmental} stray fields are from man-made objects, which are not part of the accelerator. This includes stray fields from the electrical grid, such as power lines and power stations, and nearby transport infrastructure, such as train and tram lines.

The electrical grid is typically the largest stray field source. In Europe, the electrical grid operates at 50\,Hz. This motivates the choice of 50\,Hz for the repetition frequency for CLIC. Stray fields at 50\,Hz have the same impact on a pulse-by-pulse basis. Therefore, stray fields at 50\,Hz (and higher-order harmonics) appear as if they are static to the beam and can be removed during beam-based alignment~\cite{clic-cdr}.

\textbf{Technical} stray fields are from elements of the accelerator, e.g. magnets, RF systems, power cables, etc. These stray fields are the biggest concern because of their proximity to the beam. Measurements at live accelerator facilities, which include stray fields from technical sources, are presented in Sec.\,\ref{s:lhc-measurements}.

\subsubsection{Sensitivity}
CLIC is sensitive to extremely small stray fields down to the level of 0.1\,nT~\cite{sf-impact, edu, snuverink, sf-tolerances}. This is several orders of magnitude lower than the typical level of stay fields found in accelerator environments. Therefore, they are a serious consideration in the design and operation of CLIC.

A model that characterises a realistic amplitude for stray fields in an accelerator environment is needed to simulate the impact of stray fields on CLIC and to evaluate the effectiveness of different mitigation strategies. In this paper, we work towards developing such a model.

\section{Measurements}
The magnetic field sensors used in this work are described in Sec.\,\ref{s:sensors}. The calculation of useful quantities to characterise stray fields is described in Sec.\,\ref{s:psd-corr}.

Measurements in a realistic magnetic environment for an accelerator are presented in Sec.\,\ref{s:lhc-measurements}. These measurements were taken in the Large Hadron Collider (LHC) tunnel. The LHC~\cite{lhc-design-report} is a circular proton-proton collider that uses superconducting bending magnets. It is housed approximately 100\,m underground.

\subsection{Magnetic Field Sensors}\label{s:sensors}
Four fluxgate magnetometers (Mag-13s) produced by Bartington Instruments, UK~\cite{bartington} were used in the measurements. The key specifications of these sensors are summarised in Table~\ref{t:sensor-specs}. Further details can be found in~\cite{mag13-brochure, mag13-operation-manual}. Dedicated measurements were performed to characterise the sensors, this includes measurements of their transfer function and noise curve. These measurements can be found in~\cite{thesis}. Given that we would like to measure sub-nT magnetic fields, the sensors must have a sufficiently low noise curve. The measurements in~\cite{thesis} show that the integrated noise of the sensor and DAQ over the frequency range of the sensor is small enough to measure amplitudes of 0.1\,nT. The noise level at 1\,Hz is given in Table~\ref{t:sensor-specs}.

\begin{table}[hbt]
\centering
\begin{tabular}{l c c}
\toprule
 \textbf{Specification} & \textbf{Value} & \textbf{Unit} \\
\midrule
Frequency range & 0-3 & kHz \\
Noise level (at 1 Hz) & $<$7 & pT/$\sqrt{\textnormal{Hz}}$ \\
Resolution (24-bit DAQ) & 6 & pT \\
Magnetic field range & $\pm$100 & $\mu$T \\
\bottomrule
\end{tabular}
\caption{\small Mag-13 specifications~\cite{mag13-brochure}.}
\label{t:sensor-specs}
\end{table}

The sensors require a power supply unit (PSU)~\cite{psu1}, which is also provided by Bartington Instruments. The Mag-13 sensors output an analogue voltage. A National Instruments (NI) data acquisition system was used to digitise the signal. This was a 24-bit NI 9238 module~\cite{ni9238}. The data was recorded using a NI LabVIEW script~\cite{labview} running on a laptop~\cite{dell-laptop}.

\begin{figure}[!htb]
\centering
\includegraphics[width=0.55\linewidth]{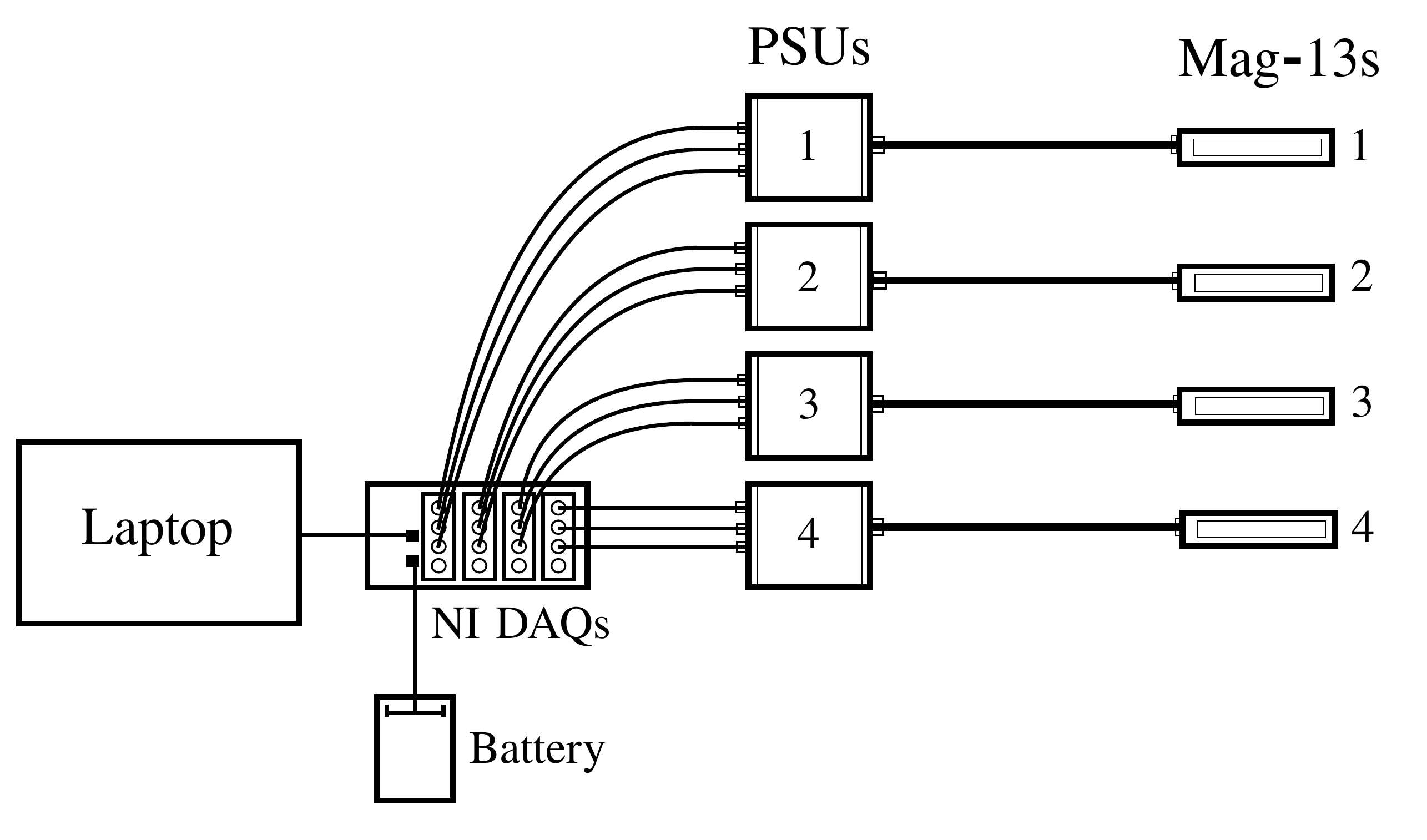}
\caption{\small Full measurement setup for survey stray fields.}
\label{f:measurement-setup}
\end{figure}

A schematic diagram of the full measurement setup is shown in Fig.\,\ref{f:measurement-setup}. All devices are powered using batteries to ensure currents from the mains do not contaminate the measurement. The setup is highly portable, which is necessary for surveying stray fields.

\subsection{Power Spectral Density and Correlation}\label{s:psd-corr}
There are two useful quantities that can be used to characterise stray fields: the power spectral density (PSD) and correlation. The PSD is the average power density as a function of frequency. This is useful for characterising the amplitude of stray fields. The correlation describes the phase difference as a function of frequency and location. This is useful for characterising the spatial variation of stray fields.

\subsubsection{PSD}
The magnetic field sensors output a voltage $v(t, s)$, which is measured as a function of time $t$ and location $s$. A periodogram can be estimated as
\begin{equation}
p_V(f, s) = \frac{1}{\Delta f} V^*(f, s) V(f, s),
\label{e:periodogram}
\end{equation}
where $V(f, s)$ is the normalised Fast Fourier Transform (FFT) of the signal $v(t,s)$, $\Delta f = f_s/N$ is the frequency bin width of the FFT, $f_s$ is the sampling frequency, $N$ is the number of data points in $v(t, s)$ and $^*$ denotes the complex conjugate.

A FFT assumes a signal is repeated infinitely many times. This often leads to discontinuities that the interface between repetitions, which causes spectral leakage. A windowing technique is applied to the voltage $v(t, s)$ to minimise spectral leakage~\cite{windowing}. In this work, we apply a Hann window~\cite{windowing} to all voltage measurements.

A FFT describes a signal in the frequency domain over the range [$-f_s/2, f_s/2$]. FFTs are conjugate symmetric functions, i.e. $V^*(-f) = V(f)$. Therefore, negative frequencies are redundant. In this paper, the FFT and PSD of a signal will only be defined for positive frequencies.

PSDs were calculated using Welch's method~\cite{welch}. Here, the signal $v(t, s)$ is split into $M$ overlapping segments. Each segment contains a 50\% overlap with its neighbours. A periodogram is calculated for each segment $p^{(m)}_V(f)$. The estimate for a PSD is calculated by averaging each periodogram,
\begin{equation}
P_V(f, s) = \frac{1}{M} \displaystyle\sum^{M}_{n=1} p^{(m)}_V(f, s).
\label{e:welch}
\end{equation}
The error on this estimate of the PSD is proportional to $1/\sqrt{M}$. The PSD of the voltage is converted into a PSD of the magnetic field by using the transfer function of the magnetometer $S(f)$, which was provided by the manufacturer. The PSD of the magnetic field is given by
\begin{equation}
P_B(f, s) = \frac{P_V(f, s)}{|S(f)|^2}.
\end{equation}
A property of a PSD is that its integral gives the variance,
\begin{equation}
\sigma^2_B (s) = \int^\infty_0 P_B(f, s) \,\text{d}f.
\label{e:variance}
\end{equation}
In this paper, we normalise PSDs such that Eq.\,\eqref{e:variance} is true. The square root of Eq.\,\eqref{e:variance} is the standard deviation. To examine the frequency content of a signal, it is useful to calculate the standard deviation as a function of frequency range,
\begin{equation}
\sigma_B(f, s) = \sqrt{\int^\infty_{f} P_B(f', s)\,\text{d}f'}.
\end{equation}

\subsubsection{Correlation}
A correlation spectrum can be calculated for two simultaneous measurements at different locations $v(t, s_0)$ and $v(t, s)$, where $s_0$ is a reference location. The correlation for each frequency and location is given by
\begin{equation}
C_{B}(f, s) = C_{V}(f, s) = \frac{\text{Re}\{P_{V}(f, s_0, s)\}}{\sqrt{P_V(f, s_0) P_V(f, s)}},
\end{equation}
where $P_{V}(f, s_0, s)$ is the cross spectral density of $v(t, s_0)$ and $v(t, s)$, which can be calculated using Welch's method by averaging correlograms,
\begin{equation}
p_{V}(f, s_0, s) = \frac{1}{\Delta f} V^*(f, s_0) V(f, s).
\end{equation}
The correlation is a measure of the phase difference of each frequency mode at two locations. It describes whether two signals are moving in phase or anti-phase. Signals with a phase difference of $0^\circ$ ($C_{B}(f, s) = 1$) are said to be highly correlated, signals with a phase difference of $90^\circ$ ($C_{B}(f, s)=0$) or signals that vary independently are said to be uncorrelated and signals with a phase difference of $180^\circ$ ($C_{B}(f,s) = -1$) are said to be anti-correlated.

\subsection{The LHC}\label{s:lhc-measurements}
The ambient magnetic field was measured near the Compact Muon Solenoid (CMS) detector~\cite{cms}. Specifically, the measurements were taken in LSS5, which is a long straight section that precedes the detector. The measurements were taken on 29/04/2019, during long shutdown 2, over the course of one hour.

The measurements were taken at a time where accelerator elements were operational. This includes magnets, vacuum pumps, cooling, ventilation, cryogenics, lighting, etc. Of interest in this work is the stray field seen by the beam. Therefore, measurements should be taken with the sensor inside the beam pipe. However, measuring inside the beam pipe is impractical due to the limited space and access. Accurately positioning and moving the sensors inside a beam pipe is also difficult. The measurements presented in this section were taken outside of the beam pipe. All known stray field sources are located outside of the beam pipe in an accelerator.

\subsubsection{Measurement Procedure}
\begin{figure}[!htb]
\centering
\includegraphics[width=0.4\linewidth]{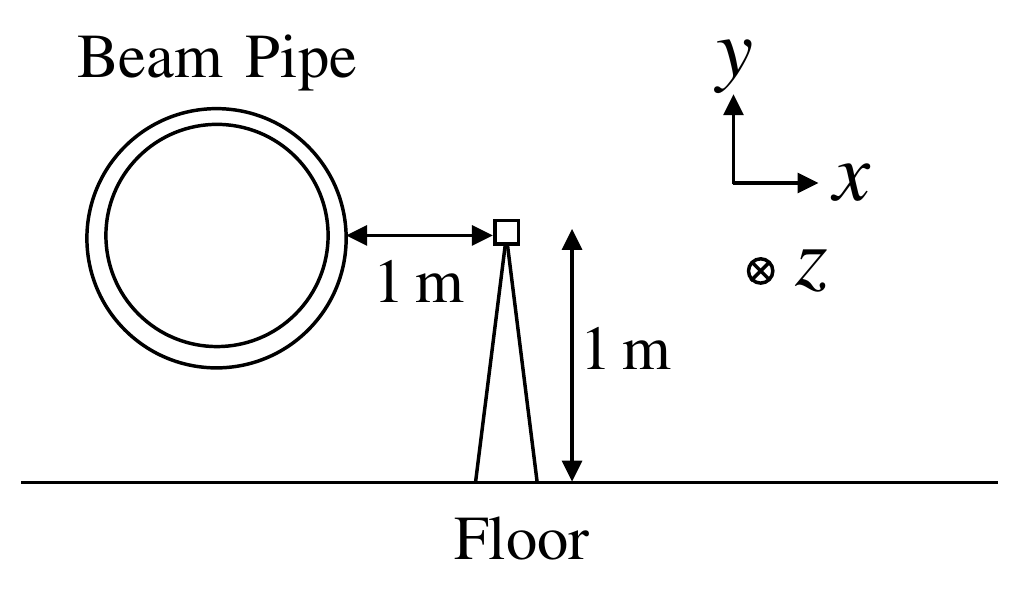}
\caption{\small Placement of the sensors relative to the beam pipe.}
\label{f:sensor-placement}
\end{figure}

Four sensors were placed at different longitudinal positions on a parallel line adjacent to the beamline (see Fig.\,\ref{f:sensor-placement}). They were approximately 1\,m away from the beamline axis. The magnetic field in three orthogonal directions: $x$, $y$ and $z$ (see Fig.\,\ref{f:sensor-placement}) was simultaneously measured for one minute by each sensor. A sampling frequency of 25\,kHz was used. Eq.\,\eqref{e:welch} was used to calculate a PSD from the measurement using $M=59$ two second segments with 50\% overlap.

In between measurements three sensors were moved to a new position along the beamline ($s$ in Fig.\,\ref{f:beamline}), mapping out a 40\,m section of the beamline at intervals of 1\,m. The fourth sensor was kept stationary as a reference.

\subsubsection{Beamline Description}
\begin{figure}[!htb]
\centering
\includegraphics[width=0.9\linewidth]{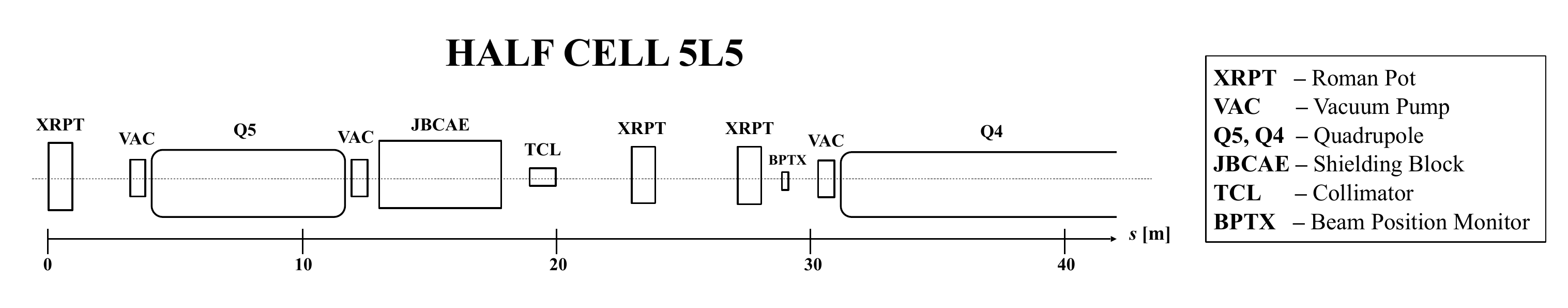}
\caption{\small Schematic diagram of the elements in LSS5. Relative lengths are to scale.}
\label{f:beamline}
\end{figure}

A schematic diagram of the elements in the beamline is shown in Fig.\,\ref{f:beamline}. The beamline includes:
\begin{itemize}
\item Two roman pots (XRPT), which are particle detectors used for machine protection~\cite{roman-pot}.
\item Three vacuum pumps (VAC), which are used to maintain the vacuum inside the beam pipe.
\item Two quadrupoles (Q5, Q4). These are the fifth and fourth closest quadrupoles to the collision point at CMS.
\item One concrete shielding block (JBCAE).
\item One collimator (TCL), which is used to collimate the beam before collision.
\item One beam position monitor (BPTX).
\end{itemize}

Objects that contain ferromagnetic materials, such as iron, affect the orientation of a magnetic field. This change in orientation is static. Elements that have a particularly high iron content are magnets, but other elements, such as the concrete block will also contain iron. Because the sensor is some distance away from the elements, we assume that the reorientation of the magnetic field will be a small effect. We will see later that an adequate description of the magnetic field measured can be generated with line currents without the need to consider iron-containing elements.

The stray fields produced by each element is a priori unknown. A measurement campaign was undertaken to characterise stray fields from particular accelerator elements. This includes magnets, rf systems, ventilation systems, vacuum pumps, etc. These measurements is presented in~\cite{thesis}. In this work, we do not aim to model the stray fields from particular sources. Instead, we will develop a model to qualitatively reproduce features in the stray field measured in the tunnel.

An important stray field source is the electrical grid. Most accelerator elements are powered from the electrical grid. We expect power cables for accelerator elements to be a large source which produces stray fields at harmonics of 50\,Hz. There is also a large amount of electrical infrastructure needed to power an accelerator, this includes power lines and sub-power stations, which will also be large stray field sources. The nearest power lines to the measurement location is at least 200\,m away from the tunnel. This is a sufficient distance to ensure that power lines are not a dominant stray field in the tunnel. Measurements of stray fields from power lines and a sub-power station on the CERN site are presented in~\cite{thesis}.

\subsubsection{PSD and Standard Deviation}
\begin{figure}[!htb]
\centering
\begin{subfigure}{.5\textwidth}
\centering
\includegraphics[width=\textwidth]{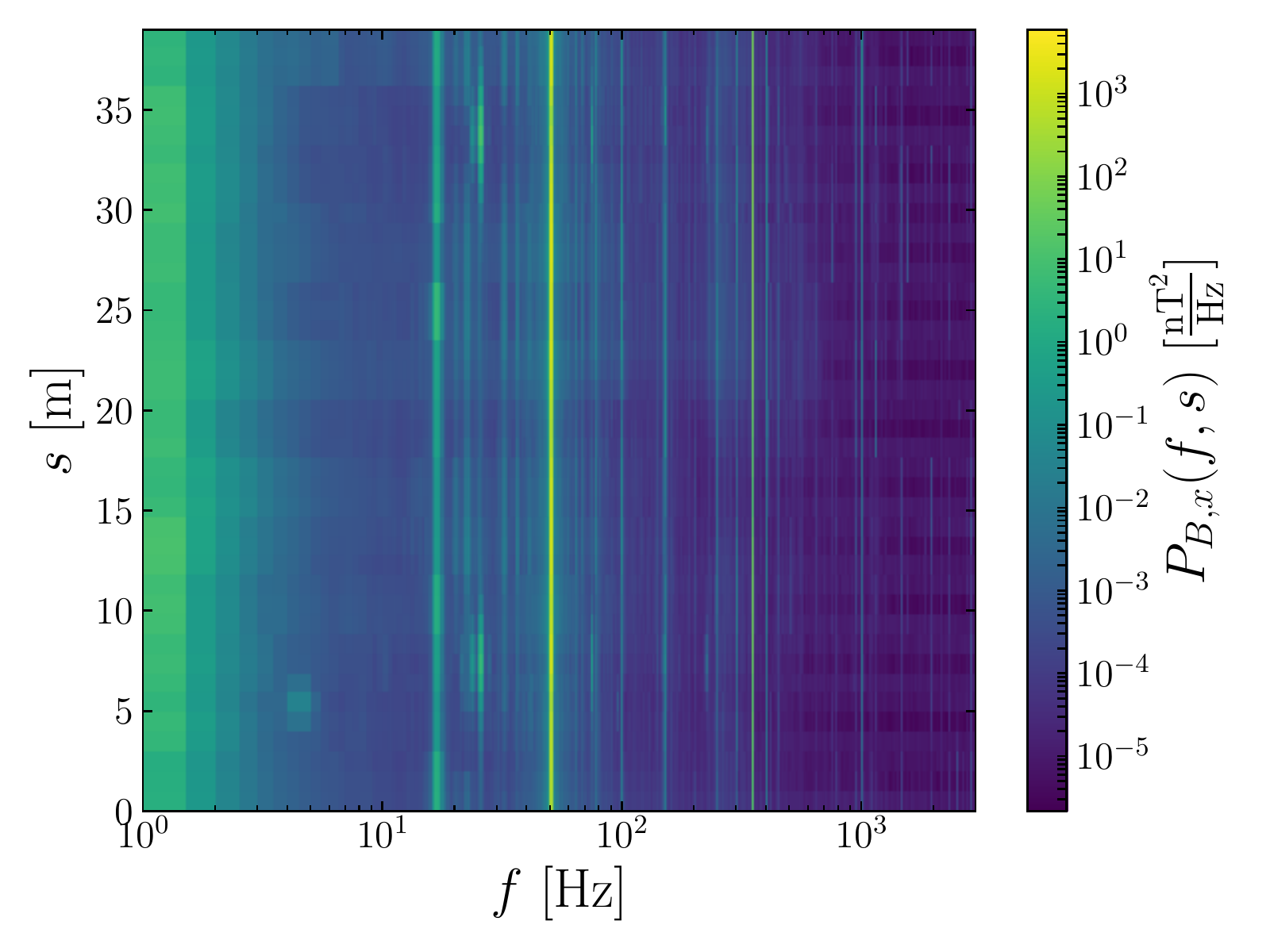}
\caption{$x$-direction.}
\end{subfigure}%
\begin{subfigure}{.5\textwidth}
\centering
\includegraphics[width=\textwidth]{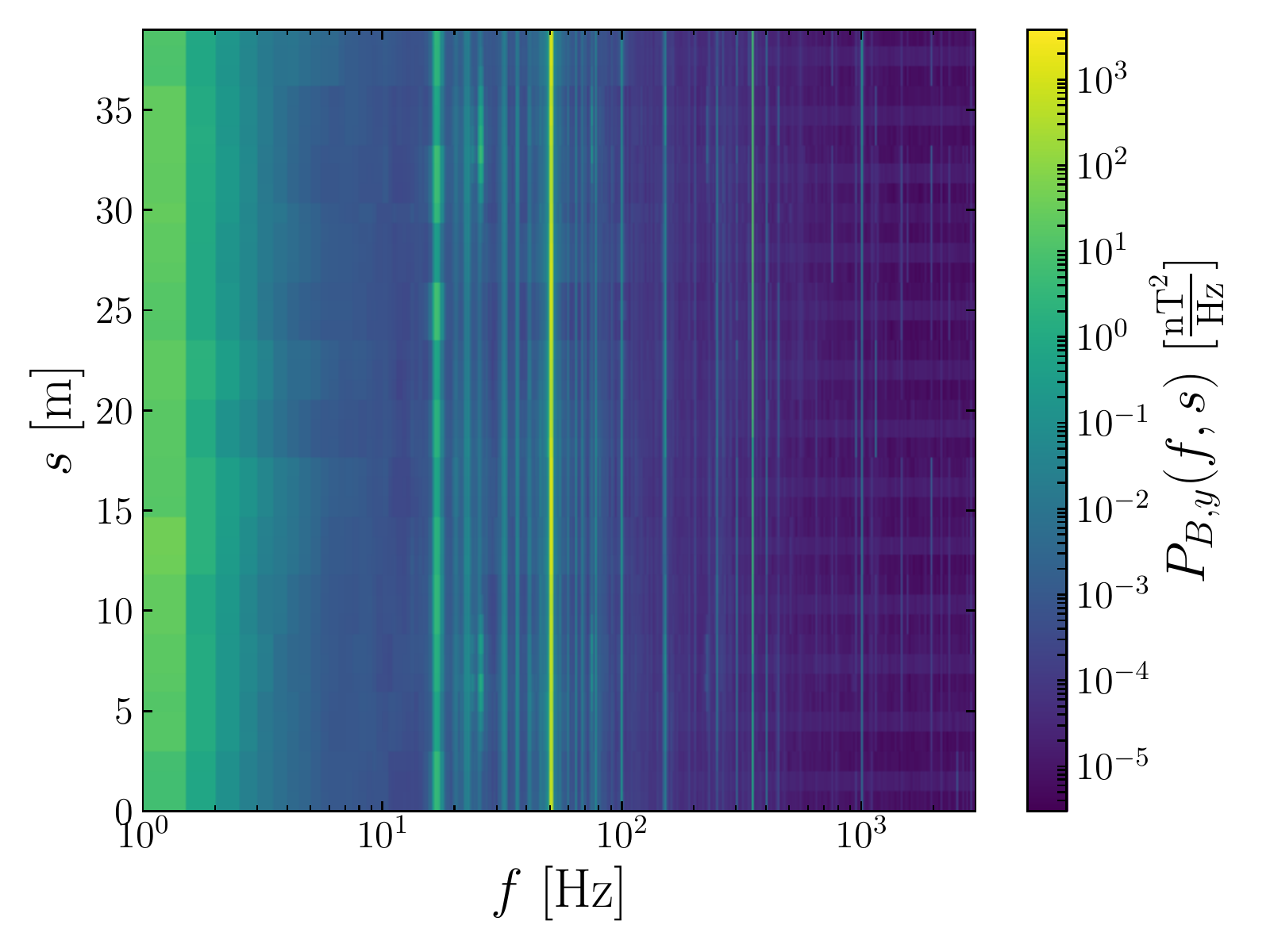}
\caption{$y$-direction. \label{f:lhc-psd-y}}
\end{subfigure}
\begin{subfigure}{.5\textwidth}
\centering
\includegraphics[width=\textwidth]{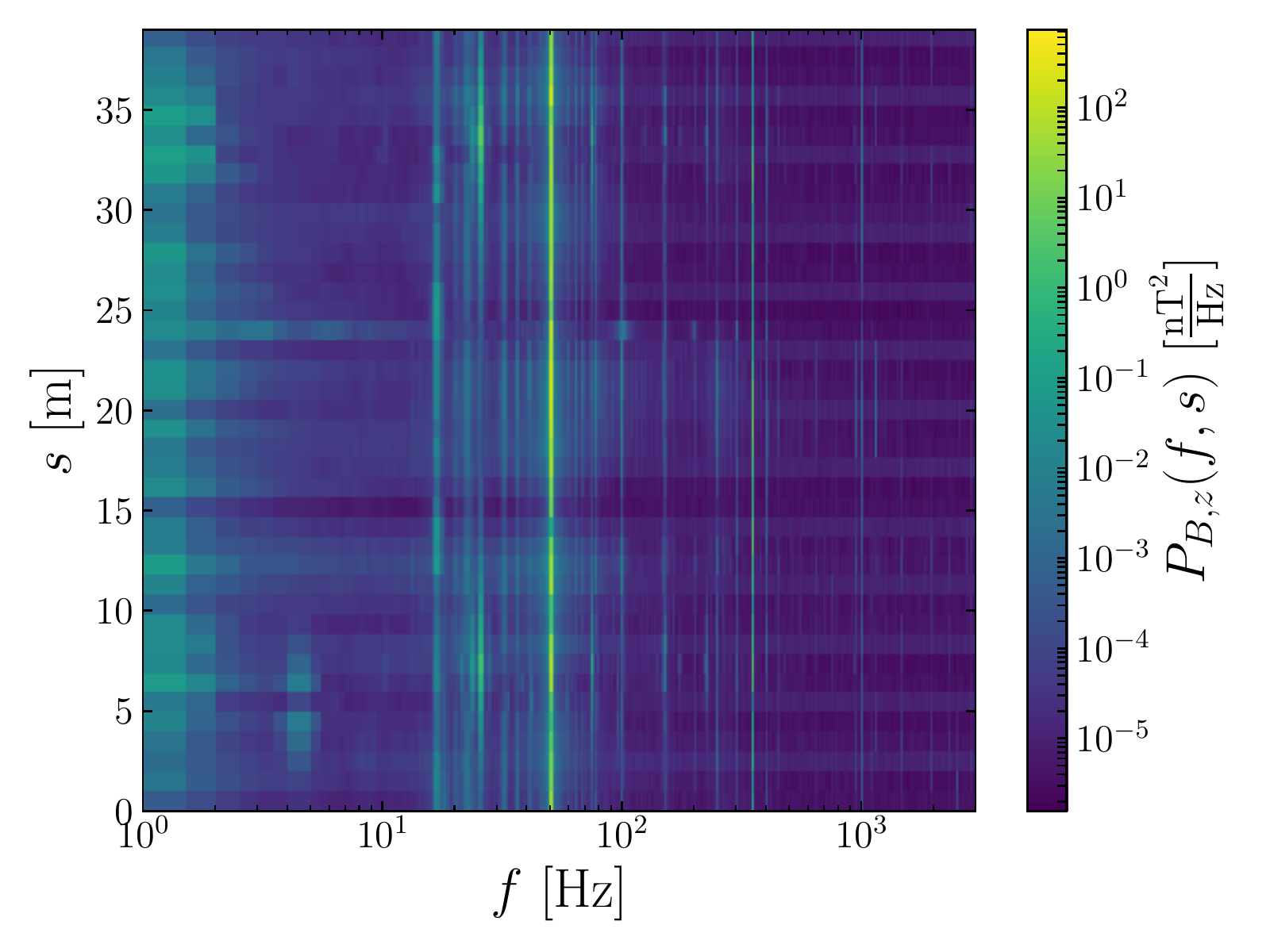}
\caption{$z$-direction.}
\end{subfigure}%
\begin{subfigure}{.5\textwidth}
\centering
\includegraphics[width=\textwidth]{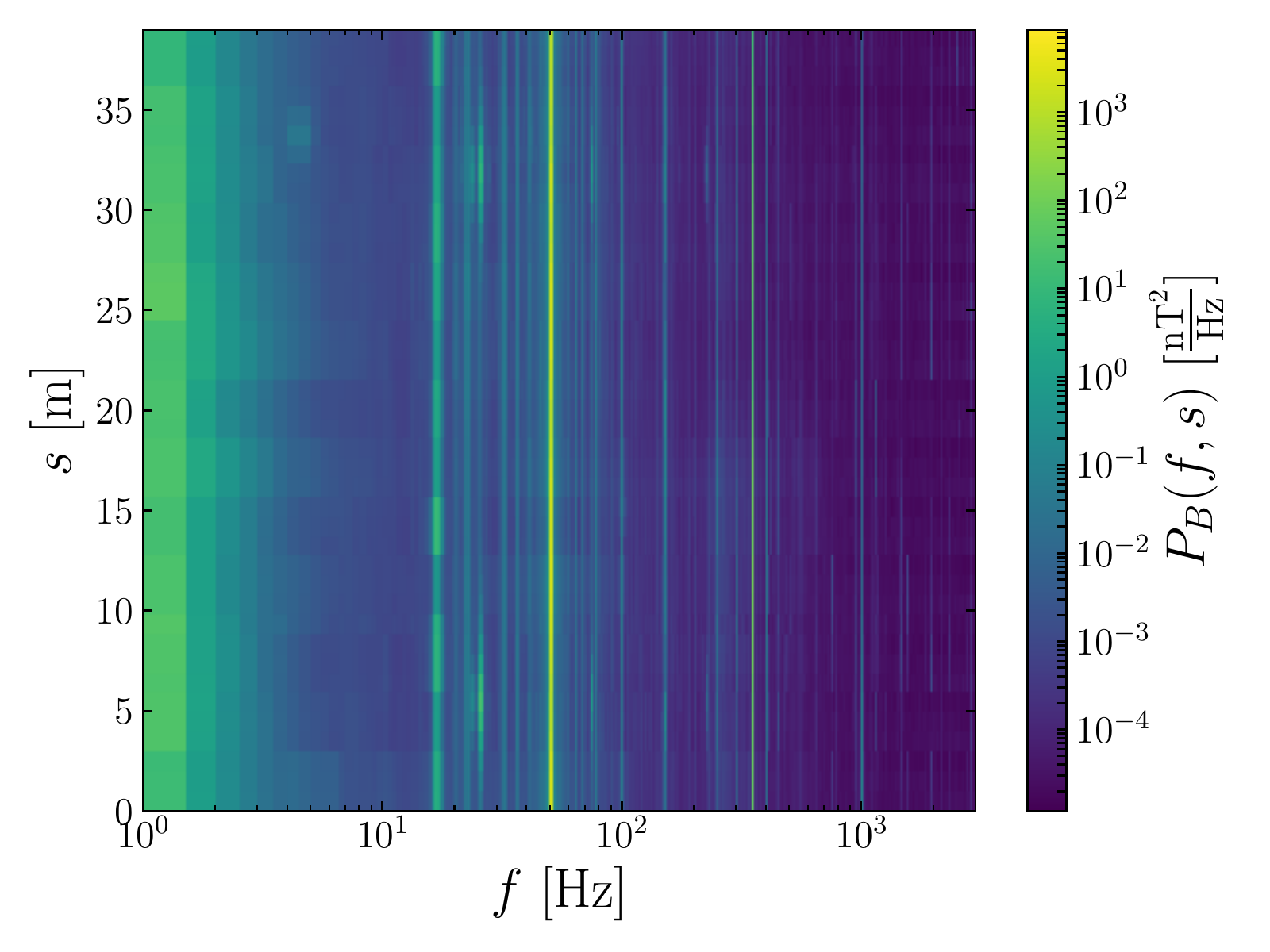}
\caption{Total. \label{f:lhc-psd-t}}
\end{subfigure}
\caption{\small PSD of the magnetic field $P_{B}(f, s)$ (RH scale) vs location $s$ (LH scale) and frequency $f$.}
\label{f:lhc-psd}
\end{figure}

\begin{figure}[!htb]
\centering
\includegraphics[width=0.55\linewidth]{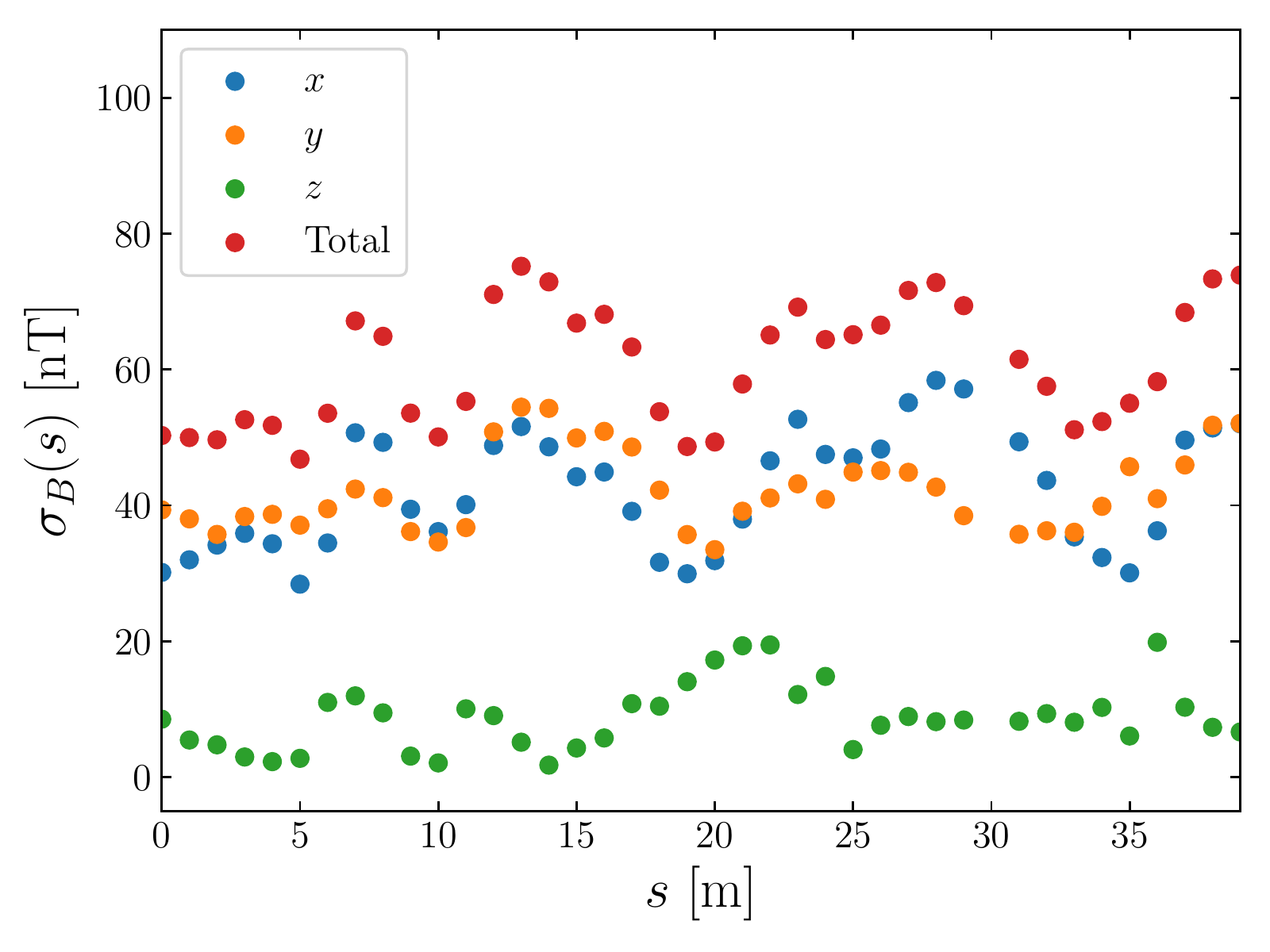}
\caption{\small Standard deviation of the magnetic field $\sigma_B(s)$ in the $x$-direction (blue), $y$-direction (orange), $z$-direction (green) and total (red) vs location $s$.}
\label{f:lhc-sd}
\end{figure}

The PSD of the magnetic field in the $x$, $y$ and $z$-direction and total PSD (sum of all three components) is shown in Fig.\,\ref{f:lhc-psd}. The amplitude of the $x$ and $y$-components is relatively constant over the length of the beamline. The $z$-component has the smallest amplitude. The most prominent peaks are at harmonics of 50\,Hz, which are from the electrical grid. The standard deviation of the magnetic field as a function of position is shown in Fig.\,\ref{f:lhc-sd}. 

\subsubsection{Correlation}
\begin{figure}[!htb]
\centering
\begin{subfigure}{.5\textwidth}
\centering
\includegraphics[width=\textwidth]{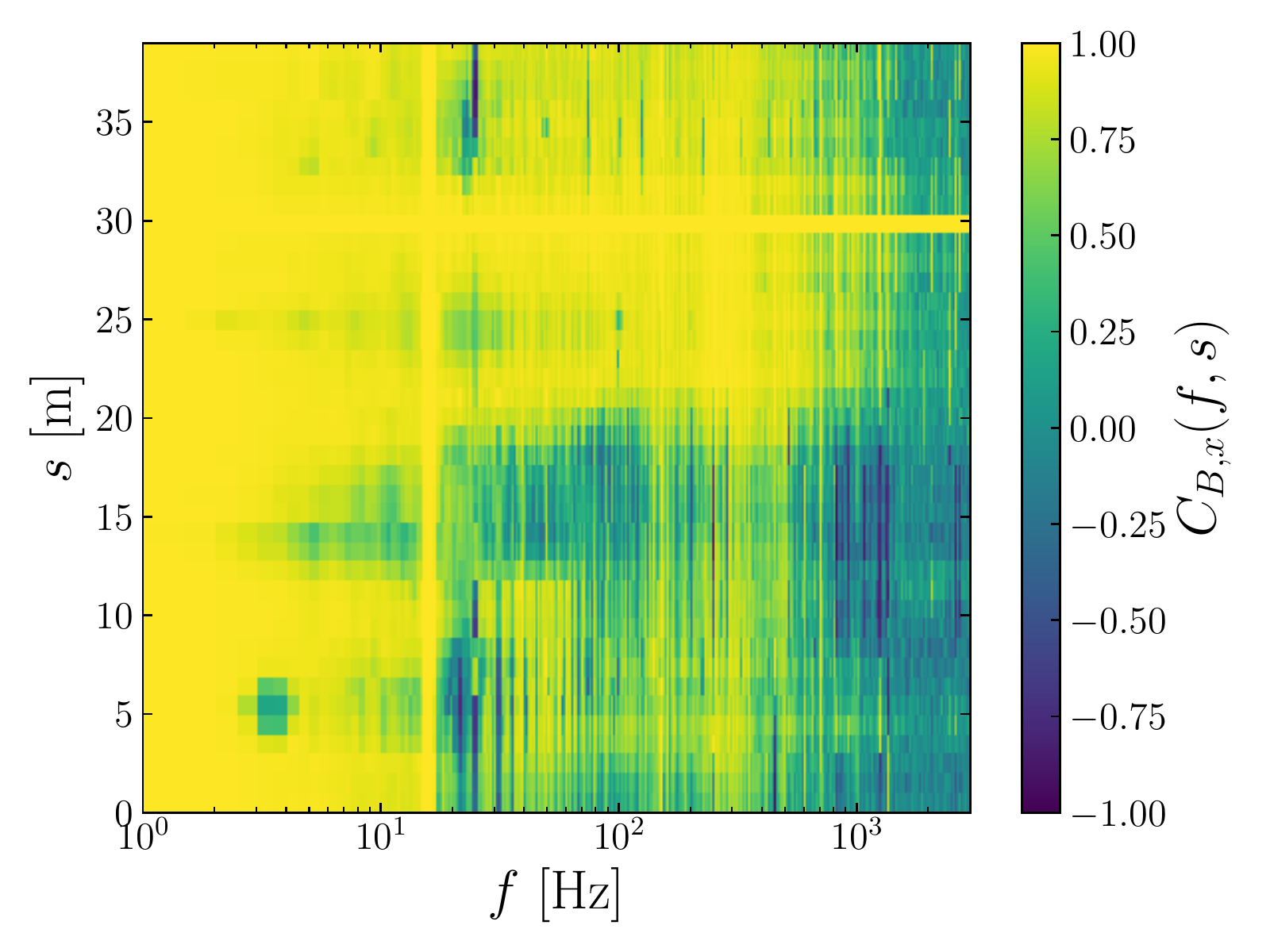}
\caption{$x$-direction.}
\end{subfigure}%
\begin{subfigure}{.5\textwidth}
\centering
\includegraphics[width=\textwidth]{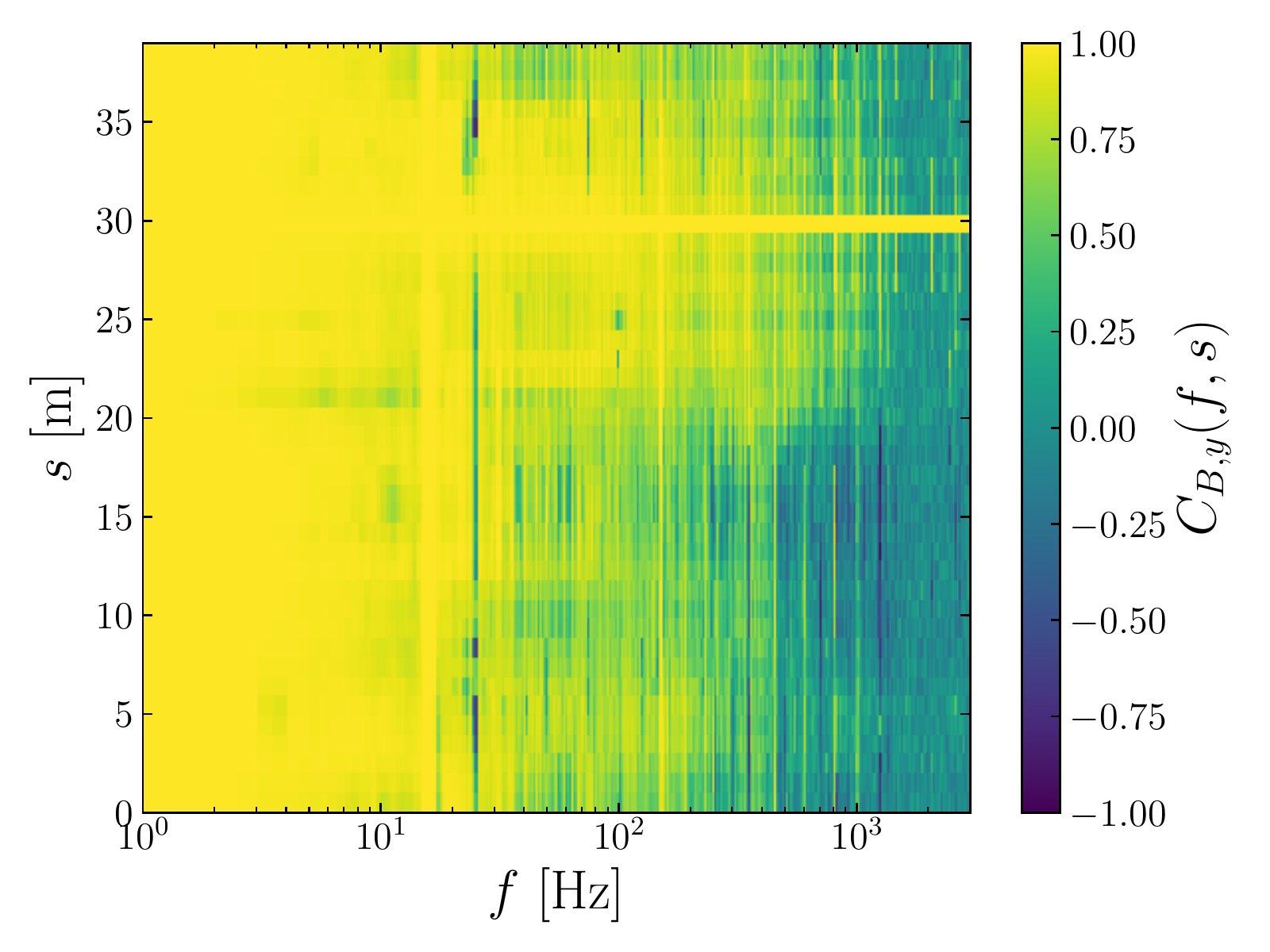}
\caption{$y$-direction. \label{f:lhc-cor-y}}
\end{subfigure}
\begin{subfigure}{.5\textwidth}
\centering
\includegraphics[width=\textwidth]{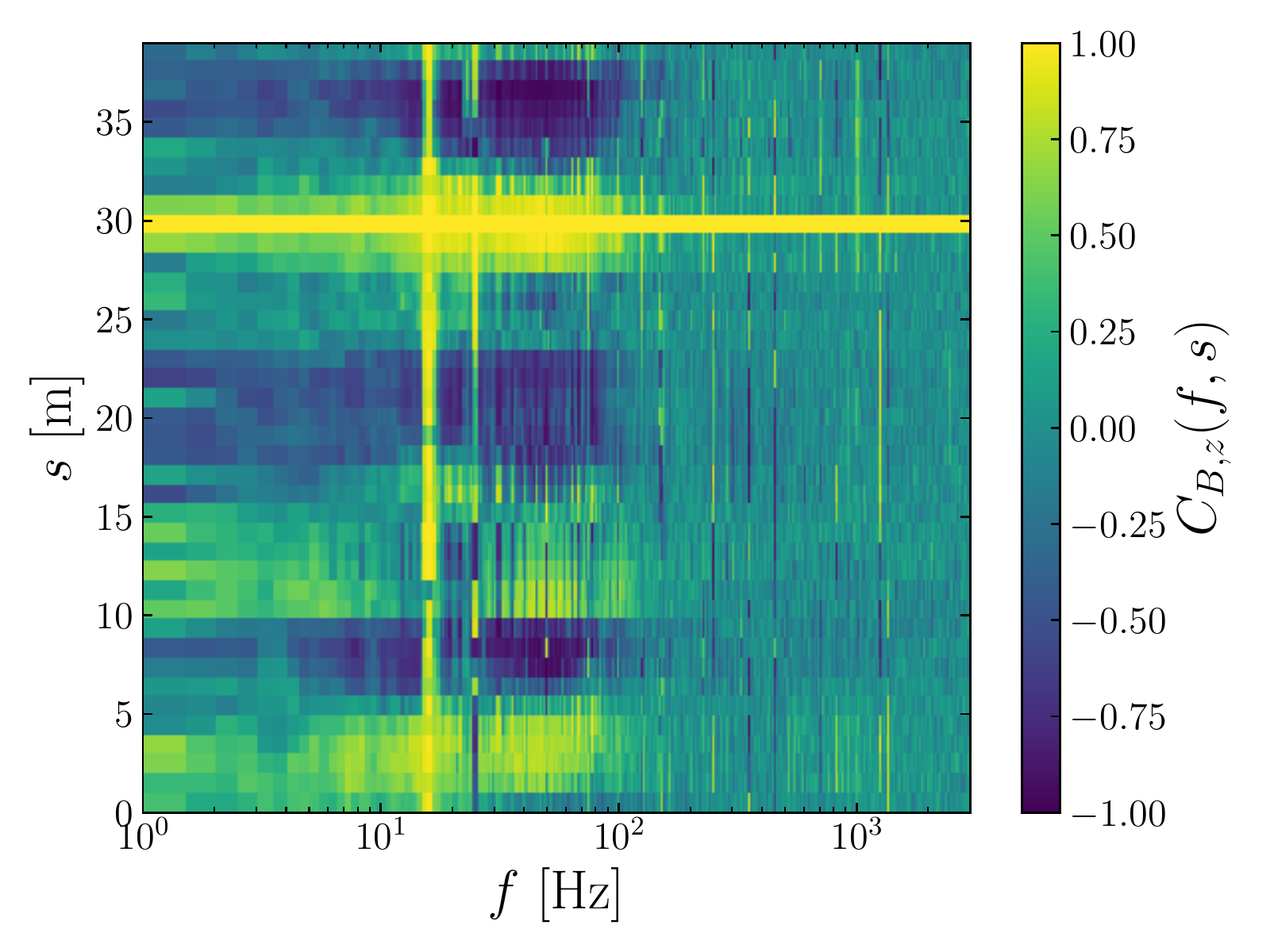}
\caption{$z$-direction.}
\end{subfigure}
\caption{\small Correlation of the magnetic field with respect to a reference sensor at $s=30$\,m $C_B(f, s)$ (RH scale) vs location $s$ (LH scale) and frequency $f$.}
\label{f:lhc-cor}
\end{figure}

The correlation of the magnetic field in the $x$, $y$ and $z$-direction with respect to the reference sensor at $s = 30$\,m is shown in Fig.\,\ref{f:lhc-cor}. The magnetic field is highly correlated for low frequencies (below 10\,Hz) in the $x$ and $y$-direction. In the $z$-direction, the magnetic field flips direction several times. The locations of anti-correlated magnetic fields coincide with the minima of standard deviation shown in Fig.\,\ref{f:lhc-sd}.

\subsubsection{Orientation of Sources}
We have observed the amplitude of the $x$ and $y$-component of the magnetic field is roughly equal and the amplitude of the $z$-component is much smaller. This is consistent with the magnetic field source being a line current in the $z$-direction. In the LHC, power cables for equipment in the tunnel run parallel to the beamline. These power cables are likely to be the dominant stray field source. 

If the magnetic field at each location originated from a common source, e.g. a single power cable parallel to the beamline, we would expect unit correlation at all locations for all frequencies. However, we observe that different frequencies have a different correlation length. In the tunnel there are a large number of cables that run parallel to the beamline. These cables can carry currents with different frequency spectra. Also not all the cables may run for the full 40\,m section fo the beamline. We believe this can lead to the correlation spectra shown in Fig.\,\ref{f:lhc-cor}. Our description of the magnetic field source is a series of many localised line current sources in the $z$-direction. In this description, we would expect that the magnetic field is highly correlated if we measure the magnetic field close to a reference location and uncorrelated if we are far from the reference location. Because the localised sources are distributed along the entire length of the beamline the reference location is not important, the correlation will only depend the separation.

The correlation in the $z$-direction does not show the low frequencies (less than 10\,Hz) to be correlated, whereas low frequencies in the $x$ and $y$-direction are. We believe this could be due to line currents oriented in the transverse plane that have a much smaller amplitude compared to the line current in the $z$-direction. This would lead to a high correlation in the $x$ and $y$-direction, which is dominated by the line current in the $z$-direction, and a correlation in the $z$-direction that is determined by the distribution of the line currents in the transverse plane. The correlation measured in the $z$-direction is consistent with transverse line currents located at $s\simeq 10$, 25 and 35\,m. Effectively, the magnetic field in the $z$-direction is due to an independent source to the magnetic field in the $x$ and $y$-direction.

\section{Modelling}
This section develops a two-dimensional PSD model for stray fields based on the LHC measurements. There are two characteristics of stray fields that must be accurately captured in the model: the amplitude and the spatial correlation.

In this paper, we follow the same approach used to simulate ground motion in linear colliders described in~\cite{seryi}. Ground motion is modelled as a set of travelling waves of differing wavenumber $k$ and frequency $f$. The amplitude of each wave is determined by a two-dimensional PSD $P(f, k)$ as
\begin{equation}
a_{ij} =  \sqrt{2} \sigma_{ij} = \sqrt{2\int^{f_{i+1}}_{f_i} \int^{k_{j+1}}_{k_j} P(f, k) \,\text{d}k \, \text{d}f} \approx \sqrt{2P(f_i, k_j) \Delta k \Delta f}.
\label{e:amplitude}
\end{equation}
The displacement of an accelerator element at a particular location and time is calculated from the superposition of each wave.

The beam is not sensitive to magnetic fields that are parallel to it. Therefore, we only need to model magnetic fields in the transverse plane ($x$ and $y$-direction in Fig.\,\ref{f:sensor-placement}).

\subsection{Amplitude}
The amplitude of the magnetic field measured in the LHC tunnel was similar in the two transverse directions to the beam ($x$ and $y$ in Fig.\,\ref{f:lhc-psd}). The $y$-component measurements were used to develop the model.

\begin{figure}[!htb]
\centering
\begin{subfigure}{.5\textwidth}
\centering
\includegraphics[width=\textwidth]{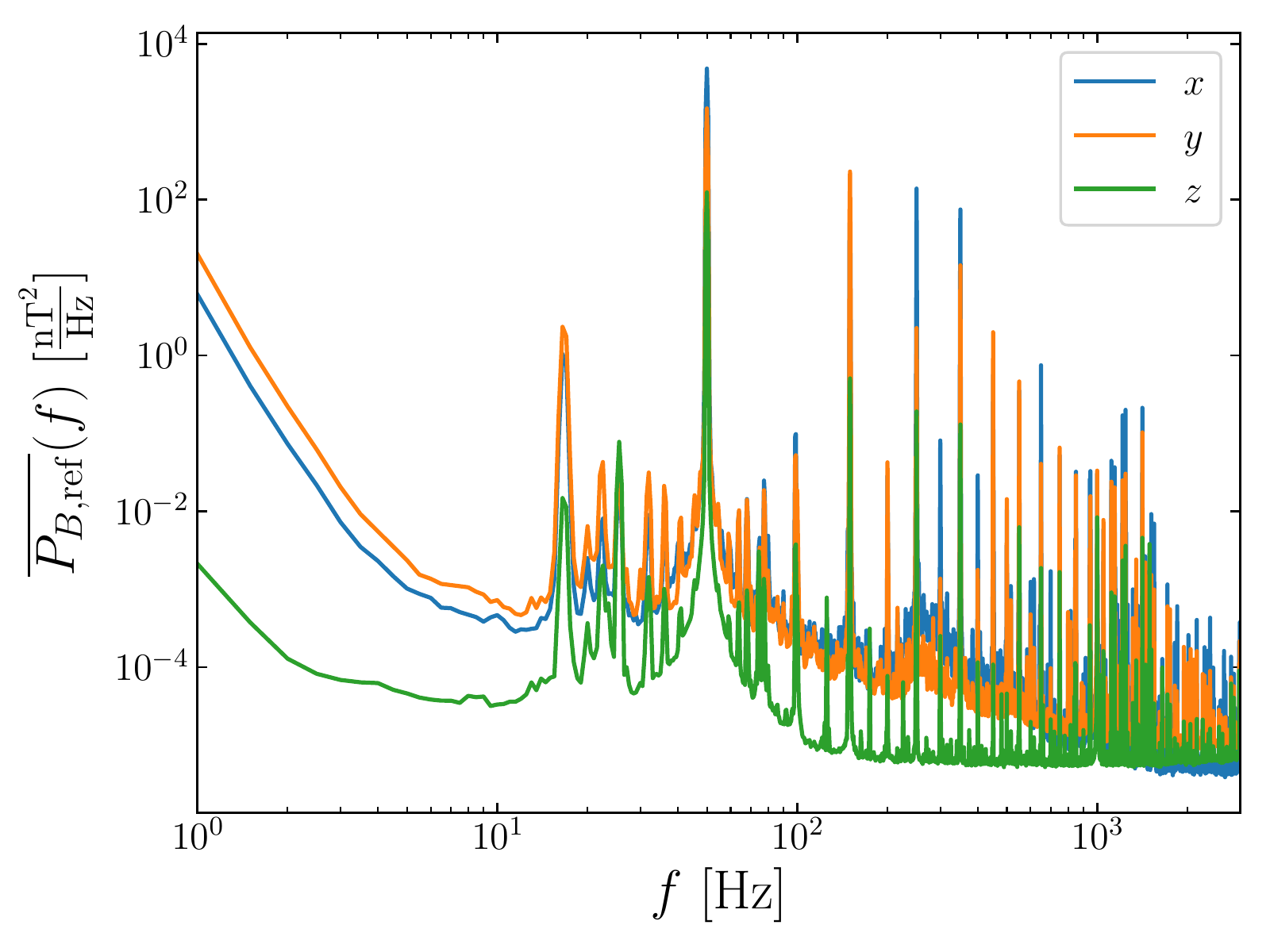}
\caption{\label{f:sf-psd-a}}
\end{subfigure}%
\begin{subfigure}{.5\textwidth}
\centering
\includegraphics[width=\textwidth]{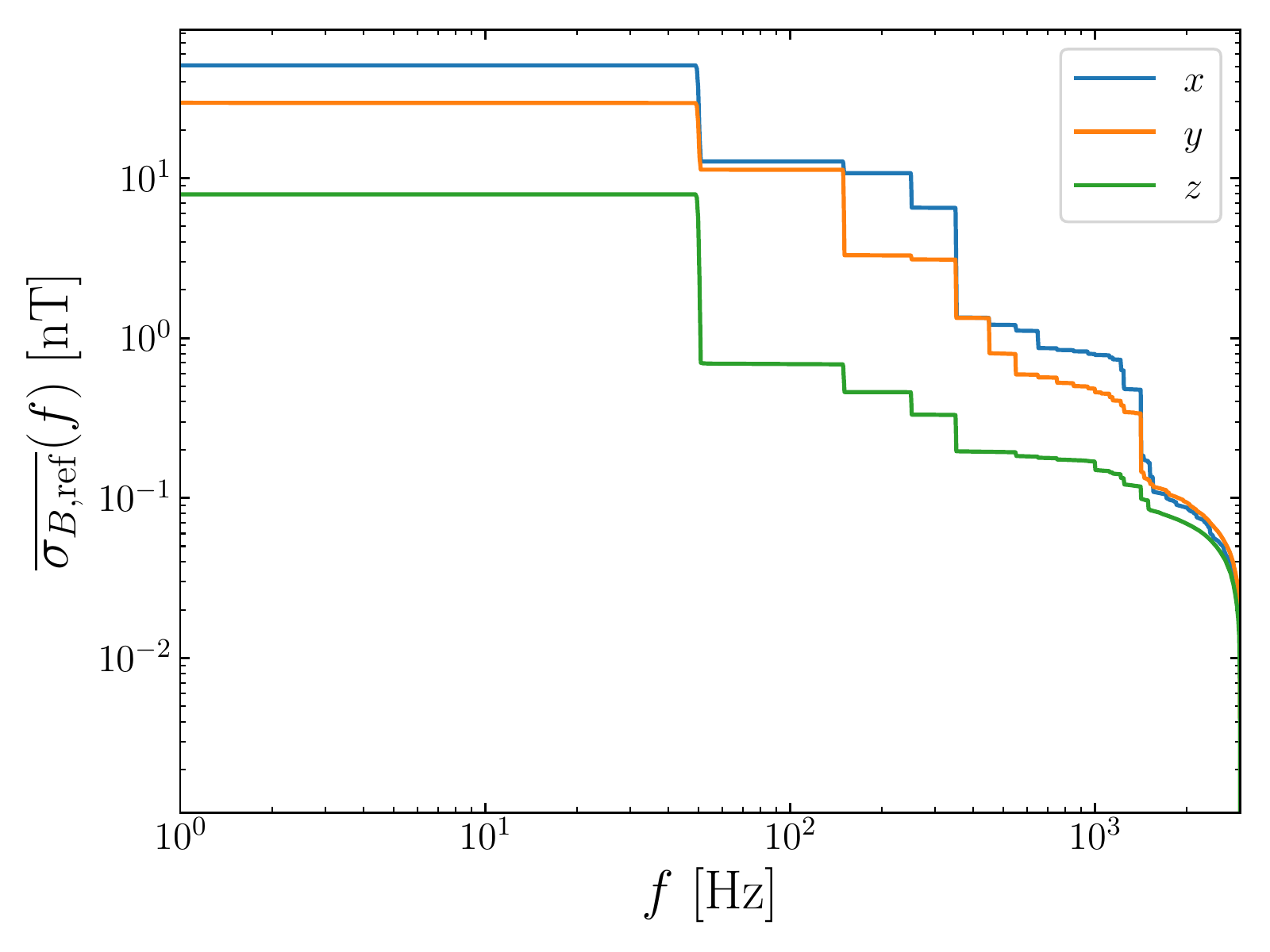}
\caption{}
\end{subfigure}
\caption{\small (a) Stray field PSD $\overline{P_{B,\text{ref}}}(f)$ vs frequency $f$ and (b) standard deviation $\overline{\sigma_{B,\text{ref}}}(f)$ vs frequency $f$.}
\label{f:sf-psd}
\end{figure}

The average PSD of the magnetic field measured by the reference sensor is given by
\begin{equation}
\overline{P_{B,\text{ref}}}(f) = \frac{1}{Q} \displaystyle\sum^Q_{i=1} P_{B,i}(f, s_\text{ref}),
\end{equation}
where $P_{B,i}(f,s_\text{ref})$ is the PSD of the magnetic field in the $x$, $y$ or $z$-direction of the $i^\text{th}$ measurement made by the reference sensor at $s_\text{ref}$ and $Q$ is the number of measurements. This is shown for each component of the magnetic field, along with the standard deviation, in Fig.\,\ref{f:sf-psd}. The $x$ and $y$-components of the magnetic field are similar. The $z$-component is approximately an order of magnitude smaller than the $x$ and $y$. There are differences in the amplitude of the 50\,Hz harmonics. This is suspected to be due to different localised sources connected to the electrical grid with different non-linear loads, which cause the harmonics to appear in the PSD.

Fig.\,\ref{f:lhc-psd} shows that the amplitude is approximately constant over the measured section. Therefore, a PSD measured at one location can be representative of the amplitude across the entire section. The PSD for the $y$-component shown in Fig.\,\ref{f:sf-psd-a} will be used to characterise the PSD of stray fields. The standard deviation of the stray field in the $y$-direction is approximately 35\,nT.

\subsection{Correlation}
The stray field model should reproduce the correlation shown in Fig.\,\ref{f:lhc-cor-y}. There are three different regions in Fig.\,\ref{f:lhc-cor-y}:
\begin{itemize}
\item Frequencies below 10\,Hz, which are highly correlated over the 40\,m section.
\item Frequencies between 10\,Hz and 400\,Hz, which are correlated over length scales of 10\,m.
\item Frequencies above 400\,Hz, which are uncorrelated.
\end{itemize}

The PSD in Fig.\,\ref{f:sf-psd} characterises the power distribution over different frequencies. To calculate a two-dimensional PSD, the power in each frequency must be distributed over different wavenumbers. The distribution over wavenumbers determines the spatial correlation of the stray field. If there are many modes of differing wavenumber, their superposition leads to an uncorrelated stray field. Whereas if the modes have similar wavenumbers, the stray field is highly correlated.

Simultaneous measurements at many locations are required to determine the wavenumber spectrum. However, only a maximum of four sensors was available for measurements. This is not enough to parameterise a wavenumber spectrum from measurements.


A particular functional form for the wavenumber spectrum must be assumed. We propose a Gaussian function for simplicity and because its width is determined by a single parameter. The power density of a mode with frequency $f_i$ and wavenumber $k_j$ is given by
\begin{equation}
P_B(f_i, k_j) = P_B(f_i)\sqrt{\frac{2}{\pi\alpha^2}} \exp \left(-\frac{k_j^2}{2\alpha^2} \right),
\label{e:sf-2d-psd}
\end{equation}
where $P_B(f_i)$ the power density of frequency mode $i$ and $\alpha$ is half the width of the distribution. We use the PSD in Fig.\,\ref{f:sf-psd}a for $P_B(f_i)$. The factor $\sqrt{2/(\pi \alpha^2)}$ was introduced to ensure that the two-dimensional PSD correctly recovers the one-dimensional PSD $P_B(f)$ after integrating over all wavenumbers,
\begin{equation}
P_B(f) = \int^\infty_0 P_B(f, k) \,\text{d}k.
\end{equation}

The width $\alpha$ is parameterised from measurements to produce a desired spatial correlation. A small value for $\alpha$ produces a stray field which is correlated over large distances, whereas a large value for $\alpha$ produces a stray field which is only correlated over short distances. The following widths were found to reproduce the correlation measured in the LHC tunnel,~\cite{thesis}
\begin{equation}
\alpha =
\begin{cases}
0.002\pi ~~\text{for}~f \leq 10\,{\text{Hz}}, \\
0.04\pi ~~\,~\text{for}~10\,{\text{Hz}}<f\leq 400\,{\text{Hz}}, \\
0.5\pi ~~~~~\text{for}~f>400\,{\text{Hz}}.
\end{cases}
\label{e:alpha}
\end{equation}

In Eq.\,\eqref{e:alpha} there are discontinuities between the different frequency bands. This leads to an underestimation of the correlation length in the model. For example, frequencies between 400 and 500\,Hz are correlated over a few metres in Fig.\,\ref{f:lhc-cor-y} but the would have a correlation length of less than one metre in the model. As uncorrelated stray fields are more harmful to the beam than correlated stray fields, this represents a pessimistic model.

\subsection{Generator}
The stray field is simulated as a grid of zero length dipoles, which is inserted into the lattice. The purpose of the generator is to calculate the kick applied by each dipole. A dipole spacing of 1\,m was used in the simulation. With this dipole spacing, only wavelengths of $\lambda_\text{min}>2$\,m can be represented. This corresponds to a maximum wavenumber of $k_\text{max}=2\pi/\lambda_\text{min} = \pi$.

The stray field is modelled as a standing wave. The stray field at location $s$ and time $t$ is given by
\begin{equation}
B(s, t) = \displaystyle\sum_{i,j} a_{ij} \cos (k_j s + \theta_j) \cos(2\pi f_i t + \phi_{ij}),
\label{e:sf-accelerator-env-generator}
\end{equation}
where $a_{ij}$ is the amplitude determined by the two-dimensional PSD (see Eq.\,\eqref{e:amplitude}) and $\theta_j$ and $\phi_{ij}$ are uniformly distributed random numbers between 0 and $2\pi$. We define $s=0$\, m to be at the collision point. Although we use cosine functions in Eq.\,\eqref{e:sf-accelerator-env-generator}, the introduction of the random phases $\theta_j$ and $\phi_{ij}$ ensures that we simulate both sine-like and cosine-like stray fields.

The computational efficiency of calculating Eq.\,\eqref{e:sf-accelerator-env-generator} can be improved by calculating a time-dependent amplitude,
\begin{equation}
A_{ij}(t) = a_{ij} \cos(2\pi f_i t + \phi_{ij}),
\end{equation}
and calculating the stray field as
\begin{equation}
B(s,t) = \displaystyle\sum_{i,j} A_{ij}(t) \cos(k_j s + \theta_j).
\end{equation}
This significantly reduces the computation time because $A_{ij}(t)$ only needs to be calculated once per time step. The stray field kick applied by each dipole is calculated using
\begin{equation}
\delta\,\text{[$\mu$rad]} = \frac{c\,\text{[m/s]} \cdot B\,\text{[nT]} \cdot L\,\text{[m]}}{E\,\text{[GeV]}} \times 10^{-12},
\end{equation}
where $c$ is the speed of light, $L$ is the dipole spacing and $E$ is the beam energy.

\begin{figure}[!htb]
\centering
\begin{subfigure}{.5\textwidth}
\centering
\includegraphics[width=\textwidth]{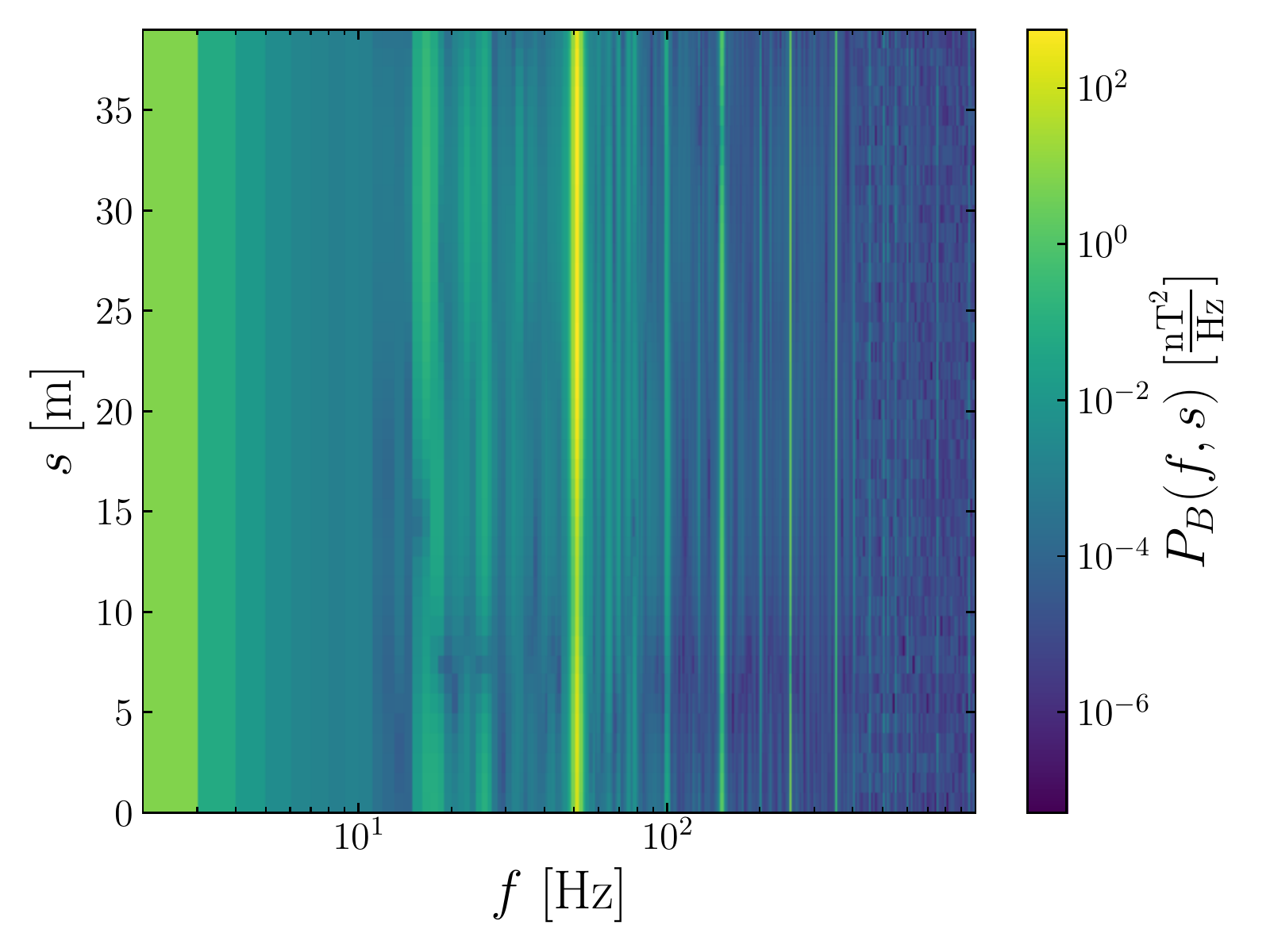}
\caption{}
\end{subfigure}%
\begin{subfigure}{.5\textwidth}
\centering
\includegraphics[width=\textwidth]{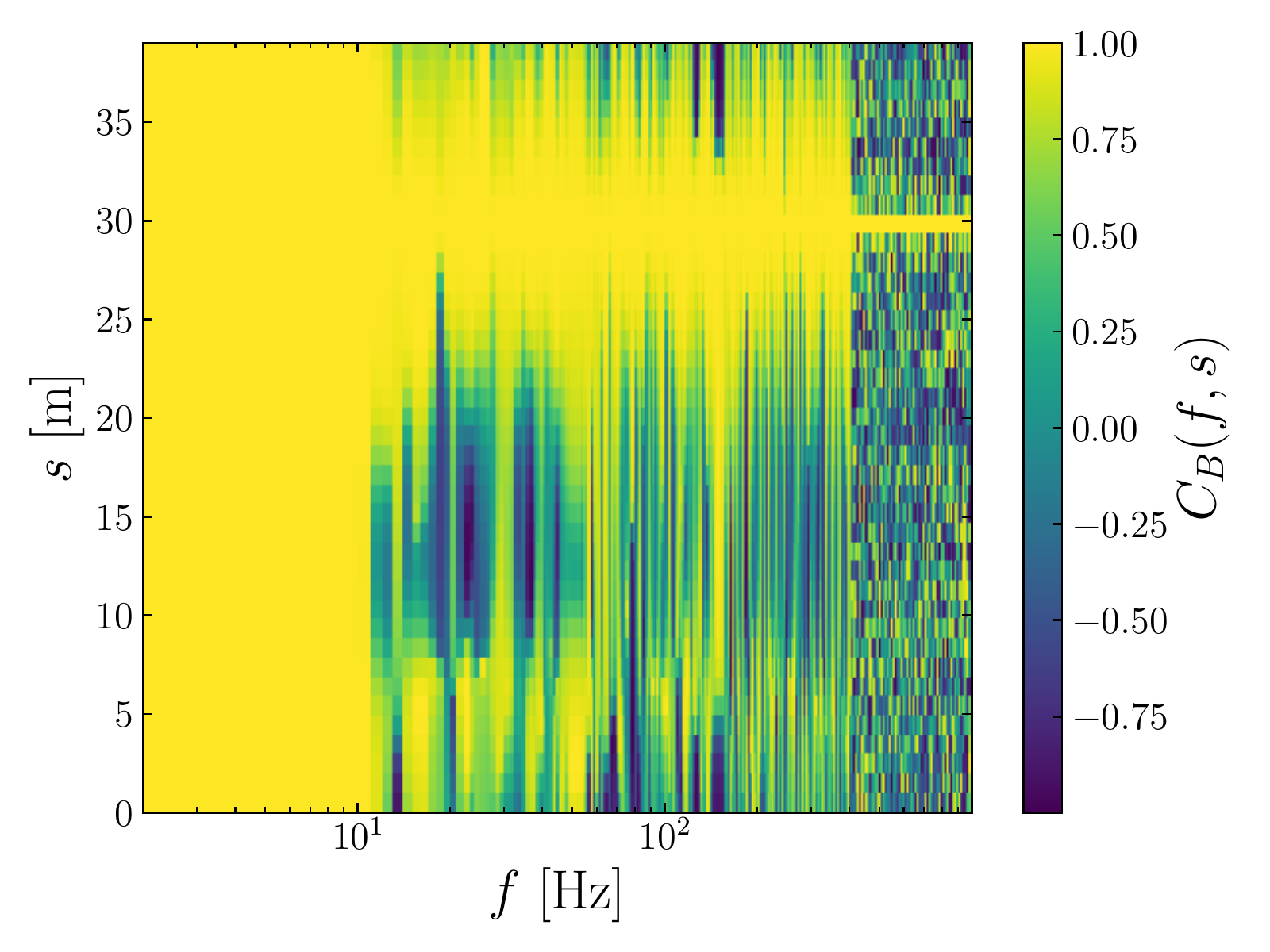}
\caption{}
\end{subfigure}
\caption{\small A sample from the generator of stray fields. (a) PSD $P_B(f,s)$ (RH scale) vs location $s$ (LH scale) and frequency $f$ and (b) correlation $C_B(f,s)$ (RH scale) vs location $s$ (LH scale) and frequency $f$. The correlation was calculated with respect to the stray field at $s=30$\,m.}
\label{f:generator-example}
\end{figure}

The generator was used to sample the stray field in a 40\,m section of the beamline. Fig.\,\ref{f:generator-example} shows the PSD and correlation of the stray field from the generator. The generator is able to qualitatively reproduce the features measured in the LHC tunnel (Figs.\,\ref{f:lhc-psd-y} and \ref{f:lhc-cor-y}).

\section{Integrated Simulations}
The two-dimensional PSD model described in the previous section was used in integrated simulations of CLIC at 380\,GeV  to evaluate the impact of stray fields on the luminosity.

\subsection{CLIC}
In this work, we combine the Ring to Main Linac (RTML), Main Linac (ML) and Beam Delivery System (BDS) of CLIC into a single tracking simulation, referred to as an `integrated simulation'. This is necessary because stray fields can be correlated over the entire length of the machine. Therefore, the entire machine must be simulated to evaluate their full effect.

The particle tracking code PLACET~\cite{placet} was used to track the electron and positron beams. A full simulation of the collision, including beam-beam effects~\cite{beam-beam-effects} was performed with GUINEA-PIG~\cite{guinea-pig} to estimate the luminosity.

\subsection{Impact of Stray Fields on the Luminosity}
For sufficiently small stray field amplitudes the time-averaged luminosity loss for a particular mode can be written as~\cite{thesis}
\begin{equation}
\left< \Delta \mathcal{L}_{ij} \right> = \frac{a^2_{ij}}{2} G^2_j = \sigma^2_{ij} G^2_j,
\end{equation}
where $\sigma^2_{ij}=a^2_{ij}/2$ is the root-mean-square amplitude, $G_j$ is the sensitivity to the wavenumber mode $j$ and $\left< . \right>$ denotes the time average. Sensitivity studies which calculate $G_j$ for stray fields in CLIC at 380\,GeV can be found in~\cite{thesis}.

Due to the temporal averaging the frequency of the stray field is not important. By summing over the frequency modes we can calculate the time-averaged luminosity loss for a particular wavenumber mode as
\begin{equation}
\left< \Delta \mathcal{L}_j \right> = \sigma^2_j G^2_j,
\label{e:lumi-loss-wavenumber-modes}
\end{equation}
where $\sigma^2_j = \displaystyle\sum_{i} \sigma^2_{ij}$. The total time-averaged luminosity loss can be calculated by summing over wavenumber modes. From Eq.\,\eqref{e:lumi-loss-wavenumber-modes} we can see that the luminosity loss is determined by root-mean-square amplitude of each wavenumber mode $\sigma_j$ and the sensitivity of the collider to that mode $G_j$. The root-mean-square wavenumber amplitude $\sigma_j$ is determined by the two-dimensional PSD [Eq.\,\eqref{e:amplitude}], whereas the sensitivity $G_j$ is determined by the accelerator lattice and beam dynamics of the collider.

The $\alpha$-parameter in Eq.\,\eqref{e:sf-2d-psd} determines the root-mean-square amplitude, or equivalently the power, of each wavenumber mode. A smaller $\alpha$ concentrates the power into a small number of wavenumbers modes and leads to a long correlation length, whereas a larger $\alpha$ distributes the power into a large number of wavenumber modes and leads to a short correlation length. The impact on the luminosity is determined by both the power in each wavenumber mode and the sensitivity of the collider to that wavenumber mode [Eq.\,\eqref{e:lumi-loss-wavenumber-modes}]. It was shown in~\cite{thesis} that CLIC has certain wavenumber modes that it is particularly sensitive to. Wavenumbers with a high sensitivity are those that match the betatron motion of the beam. Generally, increasing the $\alpha$-parameter will lead to a larger luminosity loss due to an increase in the power of wavenumber modes that are particularly harmful to the beam, whereas decreasing the $\alpha$-parameter will reduce the power in harmful wavenumber modes and lead to less luminosity loss.

\subsection{Mitigation of Stray Fields}
The impact of a mitigation system can be described using a transfer function $T(f)$, which acts on the two-dimensional PSD of stray fields $P_B(f, k)$ to give an effective two-dimensional PSD,
\begin{equation}
P_{B, \text{eff}}(f, k) = |T(f)|^2 P_B(f, k),
\label{e:mitigation-psd}
\end{equation}
which is used to generate the stray field. Here, the mitigation system only impacts the temporal variation of the stray field, i.e. all wavenumbers are affected in the same way. Therefore, Eq.\,\eqref{e:mitigation-psd} is true for mitigation systems that act equally across the accelerator.

In the following sections we look at the impact of two mitigations systems: a beam-based feedback system and a mu-metal shield.

\subsubsection{Beam-Based Feedback System}
\begin{figure}[!htb]
\centering
\includegraphics[width=0.55\textwidth]{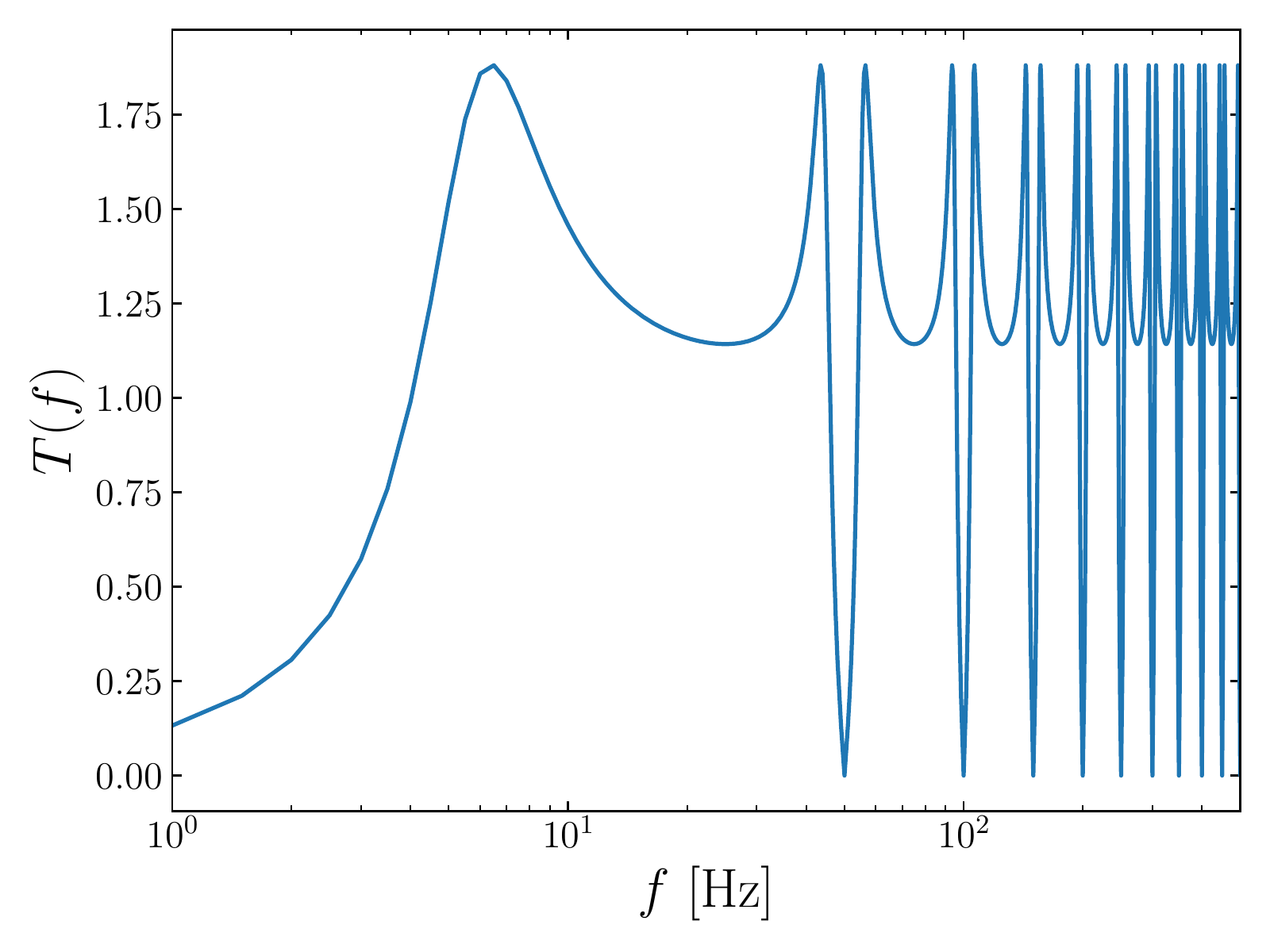}
\caption{\small Transfer function of the beam-based feedback system $T(f)$ vs frequency $f$.}
\label{f:fb-tf}
\end{figure}

The aim of the beam-based feedback system is to correct the beam offset along the accelerator. This is achieved by measuring the offset of a pulse using beam position monitors and applying a correctional kick to the following pulse using magnets. The transfer function for the CLIC feedback system is shown in Fig.\,\ref{f:fb-tf}~\cite{thesis}. The feedback system is effective at suppressing low frequency noise, below 1\,Hz, but amplifies noise in the frequency range 4-25\,Hz. The repetition frequency of the beam is 50\,Hz, which corresponds to a Nyquist frequency of 25\,Hz. Therefore, noise above 25\,Hz is aliased to lower frequencies.

This feedback system was optimised to minimise the luminosity loss from ground motion~\cite{integrated-simulations, thesis}. The same feedback system was used in stray field simulations to ensure that the proposed mitigation strategy is consistent for the combined effects of ground motion and stray fields.

\subsubsection{Mu-Metal Shield}
Another approach to mitigate stray fields is to prevent them from reaching the beam. This can be achieved by surrounding the beam pipe with a magnetic shield. The tolerances for magnetic field ripples are larger than the tolerances for stray fields~\cite{thesis}. Therefore, the magnetic shield does not need to run through the aperture of magnets, shielding the drifts is sufficient.

Ferromagnetic materials with a large magnetic permeability are commonly used to shield magnetic fields. Mu-metal offers one of the highest permeabilities. The use of mu-metal to shield magnetic fields in linear colliders is discussed in~\cite{thesis, shielding-paper}.

\begin{figure}[!htb]
\centering
\includegraphics[width=0.55\textwidth]{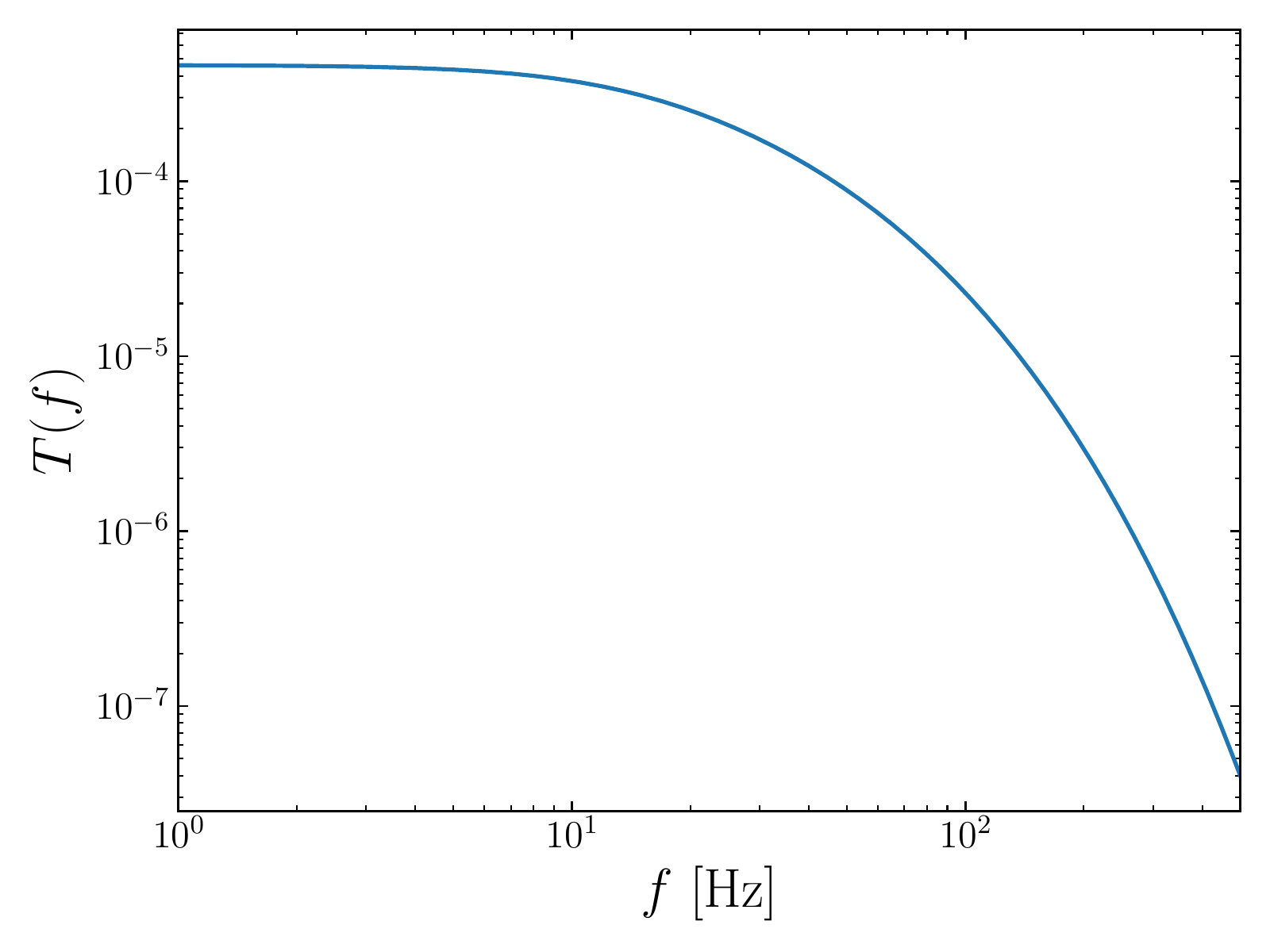}
\caption{\small Transfer function of a mu-metal cylinder with 1\,mm thickness and inner radius of 1\,cm $T(f)$ vs frequency $f$.}
\label{f:mu-tf}
\end{figure}

A methodology for calculating the transfer function of a cylindrical magnetic shield is outlined in~\cite{hoburg}. The transfer function for a cylindrical mu-metal shield with a thickness of 1\,mm and inner radius of 1\,cm is shown in Fig.\,\ref{f:mu-tf}. A relative permeability of 50,000 was used for this calculation, which is a reasonable estimate for the permeability with very low amplitude external magnetic fields~\cite{thesis, shielding-paper}. The mu-metal shield is very effective at mitigation.

Stray field simulations in~\cite{sf-tolerances, thesis} identified particular sections of CLIC, which are sensitive to stray fields. The most sensitive regions are the Vertical Transfer and Long Transfer Line in the RTML and the Energy Collimation Section and Final-Focus System in the BDS. It is possible to devise an effective mitigation strategy by just shielding these sections. The ML is the least sensitive section and benefits from shielding from the copper accelerating cavities. 

\subsubsection{Impact on the Stray Field PSD}
Ignoring the spatial variation, the transfer function can act on the one-dimensional PSD $P_B(f)$ to estimate the impact of a mitigation system at a single location. The one-dimensional PSD in Fig.\,\ref{f:sf-psd} is used to characterise the PSD of stray fields at a single location.

\begin{figure}[!htb]
\centering
\begin{subfigure}{.5\textwidth}
\centering
\includegraphics[width=\textwidth]{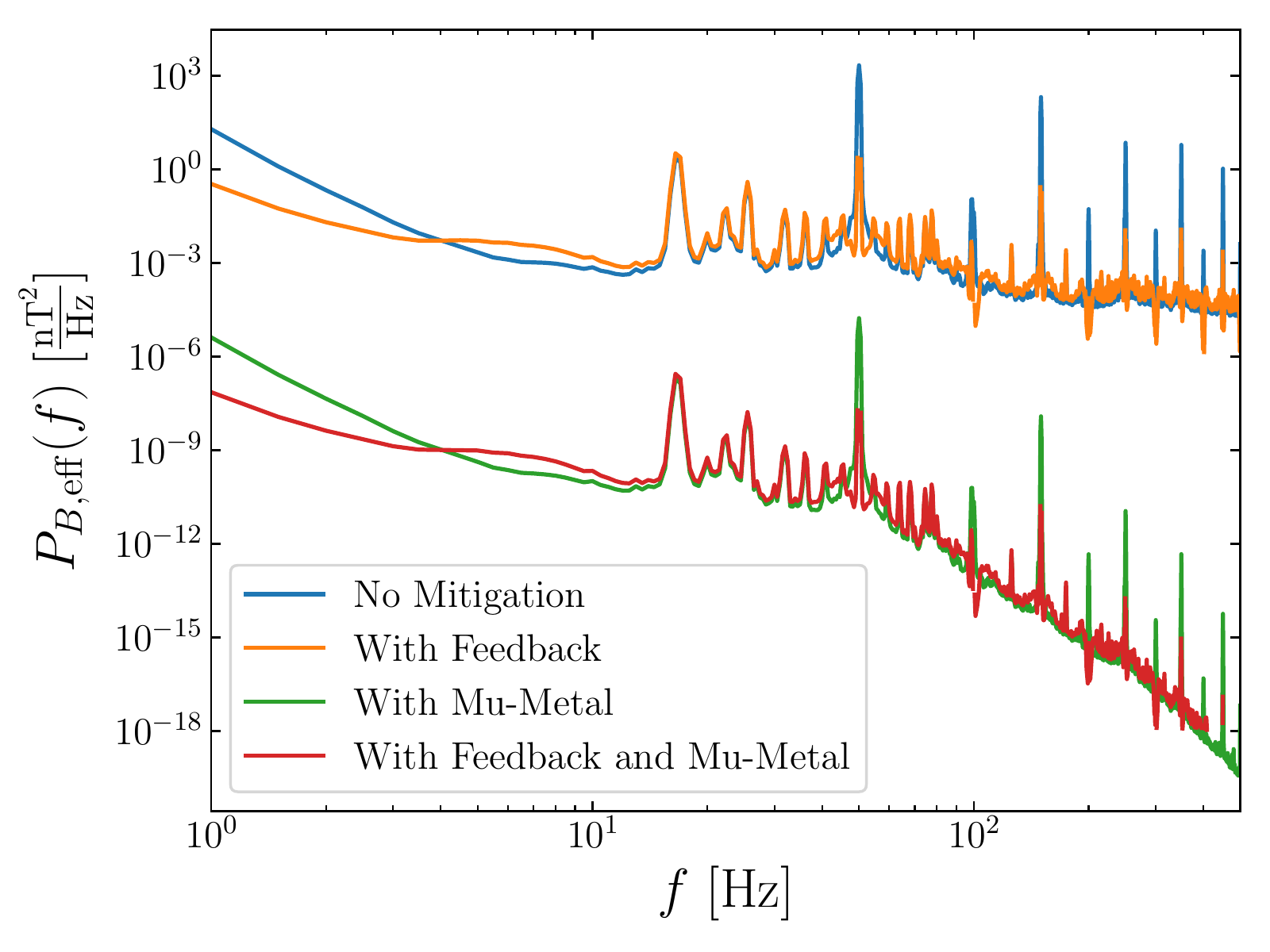}
\caption{}
\end{subfigure}%
\begin{subfigure}{.5\textwidth}
\centering
\includegraphics[width=\textwidth]{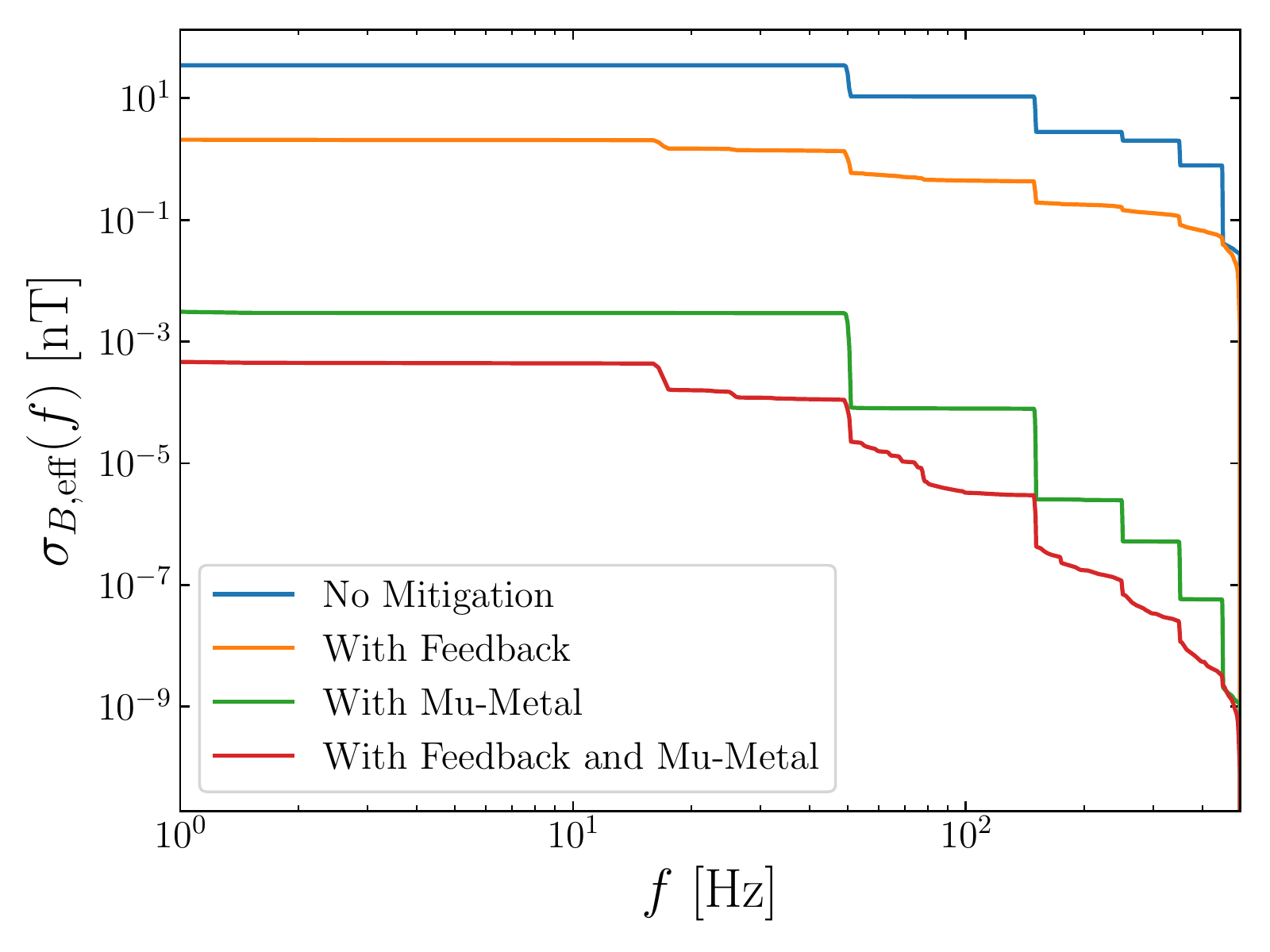}
\caption{}
\end{subfigure}
\caption{\small (a) Effective stray field PSD $P_{B,\text{eff}}(f)$ vs frequency $f$ and (b) standard deviation $\sigma_{B, \text{eff}}(f)$ vs frequency $f$: without mitigation (blue); including a beam-based feedback system (orange); including a 1\,mm mu-metal shield (green) and with the feedback system and mu-metal shield combined (red).}
\label{f:eff-sf-psd}
\end{figure}

Fig.\,\ref{f:eff-sf-psd} shows the effective PSD and standard deviation of stray fields including the impact of different mitigation systems. The standard deviations are summarised in Table~\ref{t:eff-sd}. Any stray field that is at a harmonic of 50\,Hz is cured by the beam-based feedback system. However, stray fields that aren't exactly at harmonics of 50\,Hz are not completely removed. The breadth of the 50\,Hz peak and its harmonics leads to the jumps in the standard deviation shown in Fig.\,\ref{f:eff-sf-psd}b even though a beam-based feedback system is included.

\begin{table}[!htb]
\centering
\begin{tabular}{l c}
\toprule
\textbf{Mitigation} & \textbf{$\boldsymbol{\sigma_{B,\text{eff}}}$ [nT]} \\
\midrule
None & 35 \\
Feedback System & 2.1 \\
Mu-Metal Shield & $3.1 \times 10^{-3}$ \\
Feedback System & \multirow{2}{*}{$0.5\times10^{-3}$} \\
+ Mu-Metal Shield & \\
\bottomrule
\end{tabular}
\caption{\small Standard deviation of the stray field $\sigma_{B,\text{eff}}$ with different mitigation techniques.}
\label{t:eff-sd}
\end{table}

Without mitigation, there is a large stray field of 35\,nT. With a beam-based feedback system, the effective stray field is 2.1\,nT, which is still above the 0.1\,nT level required for CLIC. The mu-metal shield is the most effective mitigation system, which brings the stray field down to the level of 3\,pT without the feedback system and 0.5\,pT with the feedback system.

\subsection{Luminosity Loss}
Integrated simulations including stray fields were performed using nominal beam parameters; additional details are provided in~\cite{thesis}. Table~\ref{t:lumi-loss} shows the luminosity loss including a beam-based feedback system and a 1\,mm mu-metal shield in sensitive regions (Vertical Transfer, Long Transfer Line, Energy Collimation Section and Final-Focus System).

Without mitigation, there is a significant luminosity loss of 43\%. The beam-based feedback system alone is not enough to mitigate stray fields. A luminosity loss of 15\% is expected if only the beam-based feedback system is used. With the mu-metal shield only, the luminosity loss is reduced to 2\%. The combination of the beam-based feedback system and mu-metal shield is an effective mitigation strategy for stray fields, reducing the luminosity loss to 0.4\%.

\begin{table}[!htb]
\centering
\begin{tabular}{l l c}
\toprule
& \textbf{Mitigation}  & $\boldsymbol{\Delta \mathcal{L}/\mathcal{L}_0}$ \textbf{[\%]} \\
\midrule
& None & 43 \\
& Feedback System & 15 \\
& Mu-Metal Shield & 2.0 \\
& Feedback System & \multirow{2}{*}{0.4} \\
& + Mu-Metal Shield & \\
\bottomrule
\end{tabular}
\caption{Relative luminosity loss $\Delta \mathcal{L}/\mathcal{L}_0$ due to stray fields. $\mathcal{L}_0$ is the nominal luminosity of CLIC.}
\label{t:lumi-loss}
\end{table}

\section{Conclusions}
High-precision magnetic field measurements were performed in the LHC tunnel, which characterised a realistic amplitude for stray fields in a live accelerator environment. These measurements were used to develop a two-dimensional PSD model, which could be used to simulate stray fields in CLIC.

The model was based on a simplified description of the sources in the LHC tunnel. The elements in the LHC tunnel are also not the same elements that would be used in CLIC. However, the LHC is on the proposed site for CLIC and at the same depth underground, which means stray fields from natural and environmental sources should be common. This work is the first step towards developing a model. To develop a more realistic model, future work is to characterise technical sources in the CLIC tunnel. Initial measurements of technical sources can be found in~\cite{thesis}.

The model for LHC-like stray fields was used to evaluate the luminosity loss that could be expected in CLIC at 380\,GeV. Integrated simulations show CLIC is robust against the level of stray fields measured in the LHC tunnel provided a beam-based feedback system and mu-metal shield is used.

\section*{Acknowledgements}
We would like to thank our CERN colleagues Benoit Salvant and Daniel Noll for their assistance in stray field measurements in the LHC tunnel.


\begin{thebibliography}{1}
\bibitem{clic-pip} M. Aicheler, et al., The
compact linear collider (CLIC)-2018 Summary report, CERN Report No. CERN-2018-010-M, 2018.
\bibitem{clic-cdr} M. Aicheler, et al., A multi-TeV linear collider based on CLIC
technology: CLIC conceptual design report, CERN Report No. CERN-2012-007, 2012.
\bibitem{sf-measurements} C. Gohil, N. Blaskovic Kraljevic, P.\,N. Burrows, B. Heilig, and D. Schulte, Measurements of stray magnetic fields at
CERN for CLIC, in \textit{Proceedings of the 10th International Particle Accelerator Conference, Melbourne, Australia, 2019}, (JACoW, Geneva, 2019), p. MOPGW081.
\bibitem{sf-impact} C. Gohil, M.\,C.\,L. Buzio, E. Marin, D. Schulte, and P.\,N. Burrows, Measurements and impact of stray fields on the 380 GeV design of CLIC, in \textit{Proceedings of the 9th International Particle Accelerator Conference, Vancouver, BC, Canada, 2018}, (JACoW, Geneva, 2018), p. THPAF047.
\bibitem{edu} E. Marin, D. Schulte, and B. Heilig, Impact of dynamical
stray fields on CLIC, in \textit{Proceedings of the 8th International Particle Accelerator Conference, Copenhagen, Denmark, 2017}, (JACoW, Geneva, 2017), p. MOPIK077.
\bibitem{balazs} B. Heilig, C. Beggan, and J. Lichtenberger, Natural sources of geomagnetic field variations, CERN Report No. CERN-ACC-2018-0033, 2018.
\bibitem{integrated-simulations} C. Gohil, D. Schulte and P.\,N. Burrows, Integrated simulation of dynamic effects for the 380 GeV CLIC design, CERN Report No. CERN-ACC-2018-0051, 2018.
\bibitem{lumi-paper} C. Gohil, P.\,N. Burrows, N. Blaskovic Kraljevic, A. Latina and D. Schulte, Luminosity performance of the compact linear collider at 380 GeV with static and dynamic imperfections, Phys. Rev. Accel. Beams \textbf{23}, 101001 (2020).
\bibitem{snuverink} J. Snuverink, W. Herr, C. Jach, J.-B. Jeanneret, D. Schulte, and F. Stulle, Impact of dynamic magnetic fields on the CLIC main beam, in \textit{Proceedings of the 1th International Particle Accelerator Conference, Kyoto, Japan, 2010}, (JACoW, Geneva, 2010), p. WEPE023.
\bibitem{sf-tolerances} C. Gohil, D. Schulte, and P.\,N. Burrows, Stray magnetic field tolerances for the 380 GeV CLIC design, CERN Report No. CERN-ACC-2018-0052, 2018.
\bibitem{lhc-design-report} O. Br\"{u}ning, S. Oliver, P. Collier, P. Lebrun, S. Myers, R. Ostojic, J. Poole, P. Proudlock, LHC design report, CERN Report No. CERN-2004-003-V-1, 2004.
\bibitem{bartington} Bartington Instruments Ltd, UK, \texttt{https://www.bartington.com}.
\bibitem{mag13-brochure} Bartington Instruments Ltd, UK, Mag-13 brochure, \texttt{https://www.bartington. com/wp-content/uploads/pdfs/datasheets/Mag-13\_DS3143.pdf}.
\bibitem{mag13-operation-manual} Bartington Instruments Ltd, UK, Mag-13 operation manual, \\ \texttt{https://www.bartington.com/wp-content/uploads/pdfs/operation\_manuals/\\Mag-13\_OM3143.pdf}.
\bibitem{thesis} C. Gohil, Dynamic imperfections in the compact linear collider, DPhil thesis, University of Oxford, 2020.
\bibitem{psu1} Bartington Instruments Ltd, UK, PSU1 brochure, \\ \texttt{https://www.bartington.com/wp-content/uploads/pdfs/datasheets/\\Magnetometer\_Power\_Supplies\_DS2520.pdf}.
\bibitem{ni9238} National Instruments, NI 9238 datasheet, \\ \texttt{http://www.ni.com/pdf/manuals/376138a\_02.pdf}.
\bibitem{labview} National Instruments, LabVIEW, \\ \texttt{https://www.ni.com/en-us/shop/labview.html}.
\bibitem{dell-laptop} DELL,
\texttt{https://www.dell.com/en-us/work/shop/dell-laptops-and-notebooks/ latitude-7480-business-laptop/spd/latitude-14-7480-laptop}.
\bibitem{windowing} K. Prabhu, \textit{Window functions and their applications in signal processing} (CRC Press, 2013).
\bibitem{welch} P. Welch, The use of fast Fourier transform for the estimation of power spectra: A method based on time averaging over short, modified periodograms, IEEE Trans. Aud. and Electr. \textbf{15}, 70-73 (1967).
\bibitem{cms} CMS collaboration, The CMS experiment at the CERN LHC, J. Instrum. \textbf{3}, S08004 (2008).
\bibitem{roman-pot} M. Oriunno, M. Deile, K. Eggert, J.-M. Lacroix, S.\,J. Mathot, E.
P. Noschis, R. Perret, E. Radermacher,  and G. Ruggiero, The Roman Pot for LHC, in \textit{Proceedings of the 10th European Particle Accelerator Conference, Edinburgh, UK, 2006}, (JACoW, Geneva, 2006), p. MOPLS013.
\bibitem{seryi} A. Seryi and O. Napoly, Influence of ground motion on the time evolution of beams in linear colliders, Phys. Rev. E \textbf{53}, 5323-5337 (1996).
\bibitem{placet} The tracking code PLACET, \\ \texttt{https://gitlab.cern.ch/clic-software/placet/blob/master/doc/placet.pdf}.
\bibitem{beam-beam-effects} D. Schulte, Beam-beam effects in linear colliders, ICFA Beam Dyn. Newslett. \textbf{69}, 237-245 (2006).
\bibitem{guinea-pig} D. Schulte, Study of electromagnetic and hadronic background in the interaction region of the TESLA collider, Ph.D. thesis, University of Hamburg, 1997.
\bibitem{shielding-paper} C. Gohil, P.\,N. Burrows, N. Blaskovic Kraljevic, D. Schulte and B. Heilig, Measurements of sub-nT dynamic magnetic field shielding with soft iron and mu-metal for use in linear colliders. To be published. Accepted in JINST.
\bibitem{hoburg} J. F. Hoburg, A computational methodology and results for quasistatic multilayered magnetic shielding, IEEE Trans. on Electr. Comp. \textbf{38}, 93-103 (1996).
\end{thebibliography}
\end{document}